%% file: manuscript-djl-tb-mwm.tex
\documentclass[12pt]{article}
\include{artsettings}
\usepackage{authblk}
\usepackage{subfigure}
\usepackage{bbm}
\usepackage{color}
\graphicspath{{figures/}}

\newcommand{\defeq}{\stackrel{\textup{def}}{=}}

\title{Mapping the Similarities of Spectra: Global and Locally-biased Approaches to SDSS Galaxy Data
}
\author[1,2]{David Lawlor}
\author[3]{Tam\'{a}s Budav\'{a}ri}
\author[4,5]{Michael W. Mahoney}
\affil[1]{Statistical and Applied Mathematical Sciences Institute}
\affil[2]{Dept of Mathematics, Duke University}
\affil[3]{Dept of Applied Mathematics and Statistics, The Johns Hopkins University}
\affil[4]{International Computer Science Institute}
\affil[5]{Dept of Statistics, University of California, Berkeley}

\begin{document}
\date{}
\maketitle

\begin{abstract}
We apply a novel spectral graph technique, that of \emph{locally-biased semi-supervised eigenvectors}, to study the diversity of galaxies.
This technique permits us to characterize empirically the natural variations in observed spectra data, and we illustrate how this approach can be used in an exploratory manner to highlight both large-scale global as well as small-scale local structure in Sloan Digital Sky Survey (SDSS) data. 
In particular, we use this method in a way that simultaneously takes into account the measurements of spectral lines as well as the continuum shape. 
Unlike Principal Component Analysis, this method does not assume that the Euclidean distance between galaxy spectra is a good global measure of similarity between all spectra, but instead it only assumes that local difference information between similar spectra is reliable.
Moreover, unlike other nonlinear dimensionality methods, this method can be used to characterize very finely both small-scale local as well as large-scale global properties of realistic noisy data. 
The power of the method is demonstrated on the SDSS Main Galaxy Sample by illustrating that  the derived embeddings of spectra carry an unprecedented amount of information. 
By using a straightforward \emph{global} or \emph{unsupervised} variant of our method, we observe that the main features correlate strongly with star formation rate and that they clearly separate active galactic nuclei. 
In addition, computed parameters of the method can be used to describe line strengths and their interdependencies. 
By using a \emph{locally-biased} or \emph{semi-supervised} variant of our method, we are able to focus on typical variations around specific objects of astronomical interest.
We present several examples illustrating that this approach can enable new discoveries in the data as well as a detailed understanding of very fine local structure that would otherwise be overwhelmed by large-scale noise and global trends in the~data.
\end{abstract}

\section{Introduction}
\label{sec:intro}

The physical properties of the Universe and the internal mechanisms of galaxies are ultimately intertwined in astronomical observations. 
Characterizing the diversity of galaxies is vital not only for understanding their evolution but also to unravel the nature of dark energy in the context of our cosmological models.
While today's large-scale spectroscopic surveys provide a plethora of data, novel data analysis methods are needed to help extract astronomical insight from these data. 

Current data analysis approaches in this area generally fall into one of two categories.
In the first category, the observed spectra are fitted by semi-analytic models, e.g., \cite{bc03}, to infer model-based parameters. 
These parameters in turn provide a model-dependent physical coordinate system with absolute scales such as age or metallicity. 
Challenges for these methods typically include systematic biases due to imperfect models as well as correlated parameters.
In the second category, one adopts a more empirical approach, where galaxies are analyzed in relation to other galaxies based on the original measurements, i.e., based on the observed spectra.  
A major challenge for these more empirical methods is the conceptual problem of how best to compare empirical spectra, e.g., which features of a spectrum are most important for identifying similarities between two spectra. 
The approach we describe in this paper falls into this second category, and it aims to address the fundamental issue of measuring similarity between galaxy spectra, both with respect to the large-scale global properties of the empirical data, as well as with respect to finer-scale local structure that might be overwhelmed by global data analysis tools that focus on global properties of the~data.

\begin{figure}
      \begin{center}
         \includegraphics[width=.40\textwidth]{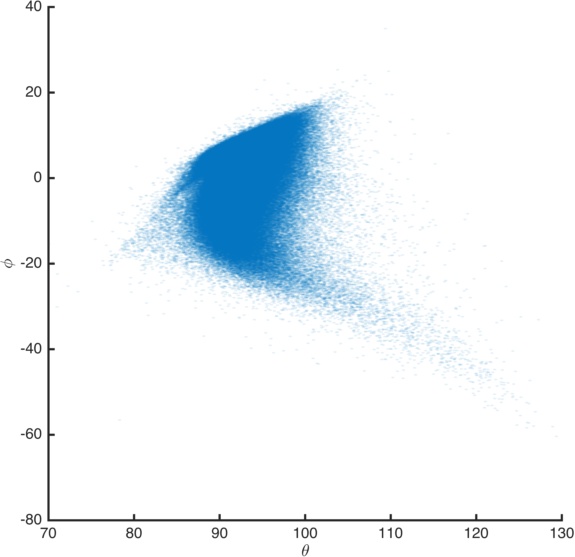}
      \end{center}
\caption{Embedding of the Main Galaxy Sample of the SDSS Data Release 7 on the mixing angles $\theta$ and $\phi$ of the first three eigen-coefficients. }
\label{fig:intro-embeddings-pca}
\end{figure}

A canonical example of a global data analysis tool that adopts this more empirical approach is Principal Component Analysis (PCA), which is widely-used to find the globally-dominant linear trends in the data.
PCA was first applied to galaxies by \cite{connolly1995spectral}, who found that a significant fraction of the variance in the spectra can be captured by only three components. 
In other words, the analyzed spectra could be well approximated by a linear combination of three \emph{eigenspectra}. 
The coefficients serve as summaries of the high-dimensional spectra, and in this coordinate system galaxies could be meaningfully compared to one another. 
PCA has been used in many research areas, including photometric redshift estimation \cite{connolly1999orthogonal,budavari2000creating}, sky subtraction \cite{wild05}, as well as classification of galaxies and quasars \cite{francis92,connolly99, yip04a, yip04b}. 
The Sloan Digital Sky Survey (SDSS) \cite{york00} has adopted the method in its data reduction pipeline, and it automatically derives the first five eigencoefficients (called \texttt{eCoeff\_0} -- \texttt{eCoeff\_4}). 
These are considered the state of the art in describing the continuum shape of the spectra. 
Figure~\ref{fig:intro-embeddings-pca} shows the mixing angles $\theta$ and $\phi$ of the three leading eigencoefficients for the Main Galaxy Sample (MGS) in SDSS Data Release 7 \cite{abazajian2009seventh}. 
These coordinates are defined as in~\cite{yip2004distributions} by
\begin{equation}
	\label{eq:theta-phi-defn}
	\phi = \tan^{-1}\left( \frac{\texttt{eCoeff\_1}}{\texttt{eCoeff\_0}} \right), \quad
	\theta = \cos^{-1}(\texttt{eCoeff\_2}).
\end{equation}
In the embedding illustrated in Figure~\ref{fig:intro-embeddings-pca}, every point is a galaxy, and nearby points (i.e., galaxies or points that are near each other in the two-dimensional representation) have similar observations, by construction of the empirical coordinate system. 
In particular, ``red and dead'' galaxies appear at the top of the plot, while star-forming blue ones are on the lower right.

There are, however, shortcomings of the simplified view of the data provided by PCA. 
First of all, a significant fraction of the galaxies are actually removed from or scattered out of the plot as a part of the usual data analysis pipeline used to generate the plot; we only really see the core of the distribution in a figure such as this one. 
In addition, the lack of structure in this visualization is surprising, especially considering the large amount of high-quality data and the wide range of galaxy types in the data. 
Finally, the interpretation of the axes is difficult, as they are linear combinations of actual data elements and not actual data elements. 
Extensions and variants of PCA have been proposed to overcome these challenges, including non-negative matrix factorization \cite{nmf}, the use of robust statistics \cite{budavari2009reliable}, and CX/CUR matrix decompositions~\cite{CUR_PNAS,yip2014objective}. 
While these methods have alleviated some of the issues associated with PCA, the fundamental limitation of its assumed linear model remained.

Perhaps the biggest conceptual change in the area was introduced by \cite{vanderplas2009reducing}, who applied the Locally Linear Embedding (LLE) method of \cite{roweis2000nonlinear}.
This more sophisticated empirical approach attempts to identify and exploit local structure in the data, and thus it broke away from the straightforward global linear model underlying PCA. 
While there are other related nonlinear approaches~\cite{tenenbaum2000global,belkin2003laplacian}, LLE in particular attempts to provide an angle-preserving mapping that assigns coordinates to galaxies such that each galaxy is approximately a linear combination of its nearest neighbors, with the same weights as in the observed space. 
The power and practical usefulness of LLE (as well as other related nonlinear methods~\cite{tenenbaum2000global,belkin2003laplacian,coifman2006diffusion}), however, is known to be severely diminished in many practical situations.
The reasons for this are many: due to the difficulty of these methods in dealing with non-uniform point densities; since the global objective function used to enforce angle preservation or other neighborhood information can damage small-scale or local structure in the data; since these methods are quite sensitive to realistic noise in the data; and since these methods are very sensitive to the ``details'' of constructing the nearest neighbor graph, e.g., to the functional form of the nearby distance and the choice of parameters used to define nearness. 
(This is in spite of a large body of theory stating that in idealized situations these details do not matter.) 
In addition to exploiting the strong algorithmic and statistical theory underlying our main method~\cite{HM14_JRNL,mahoney2012local}, dealing appropriately with these and other related practical graph construction issues will be central to our approach, and thus we postpone further discussion of it until Sections~\ref{sec:methods} and~\ref{sec:knobs}.

\begin{figure}
      \begin{center}
      \subfigure[]{
         \includegraphics[width=.30\textwidth]{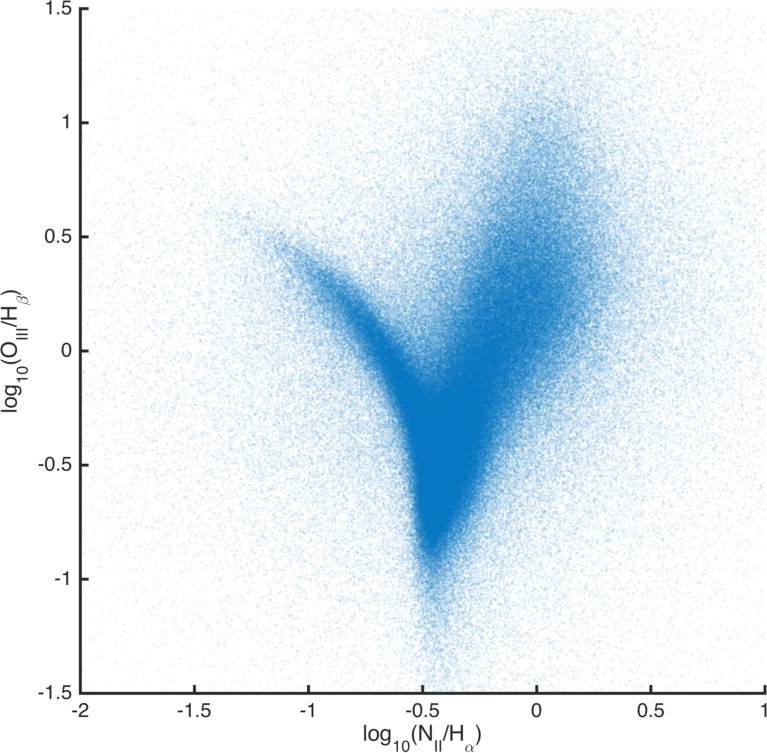}
	 \label{fig:intro-embeddings-bpt1}
      } 
      \subfigure[]{
         \includegraphics[width=.30\textwidth]{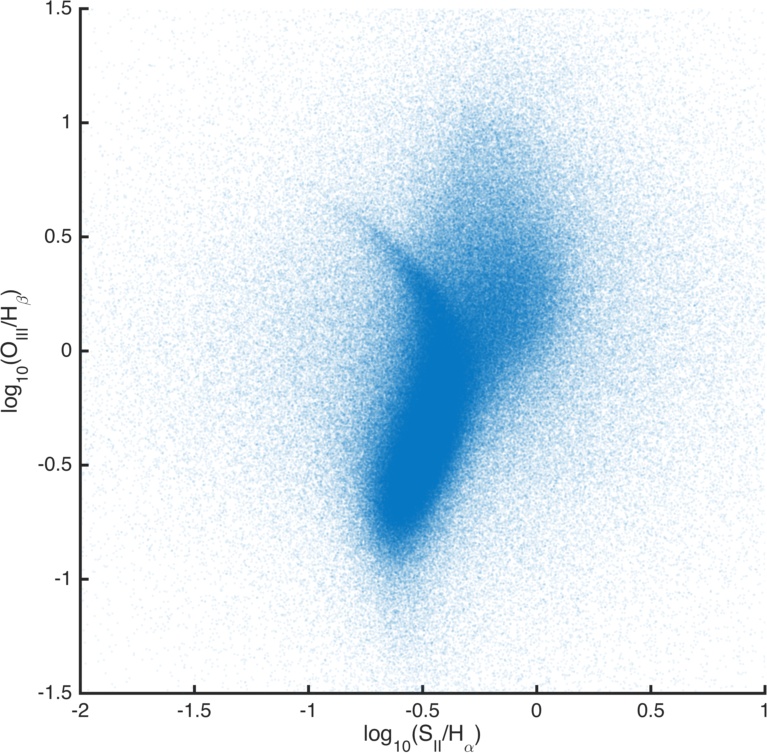}
	 \label{fig:intro-embeddings-bpt2}
      } 
      \subfigure[]{
         \includegraphics[width=.30\textwidth]{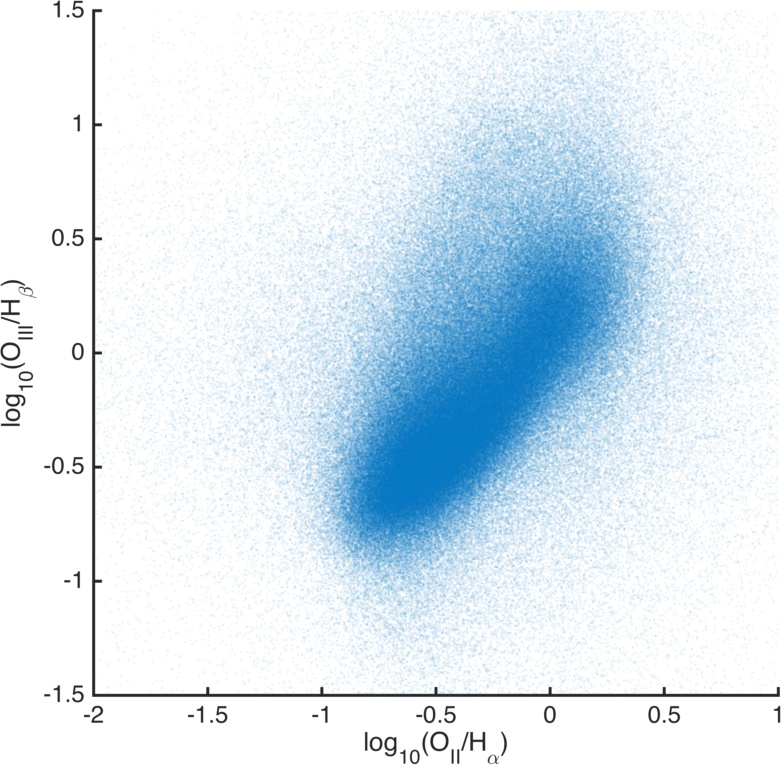}
	 \label{fig:intro-embeddings-bpt3}
      } 
      \end{center}
\caption{Embeddings based on several pairs of BPT line-ratio diagrams of the SDSS DR7 MGS.  }
\label{fig:intro-embeddings-bpt}
\end{figure}

A third empirical approach that is worth mentioning is based completely on line measurements.
Recall that the high resolution in wavelength often allows the identification and measurement of different spectral lines, and that it is common to plot spectra in terms of carefully-chosen line ratios.
That is, while not usually described as an embedding method, the typical use of line measurements often involves embedding or mapping the data to a low-dimensional space.
(The state-of-the-art in this area actually uses PCA as a pre-processing step to subtract the continuum, in order to measure better the line strength relative to this baseline~\cite{tremonti}.) 
In particular,  the BPT diagrams~\cite{baldwin1981classification} plot different line ratios on a logarithmic scale, enabling, e.g., the classification of galaxies~\cite{brinchmann04,kewley2006}. 
For example, Figure~\ref{fig:intro-embeddings-bpt} shows several BPT diagrams of the SDSS MGS. 
In Figure~\ref{fig:intro-embeddings-bpt1}, the characteristic V-shape of the embedding on the ratios $\textrm{N}_\textrm{II}/\textrm{H}_\alpha$ vs.~$\textrm{O}_\textrm{III}/\textrm{H}_\beta$ is clearly visible, despite significant scatter that is partly due to noisy measurements of the individual lines. 
Little to no structure is evident in Figure~\ref{fig:intro-embeddings-bpt2} and~\ref{fig:intro-embeddings-bpt3}, whose $x$-axes plot different line strengths. 
The insight conveyed by the BPT plots can be considered complementary to that of the PCA results, which is primary based on the continuum shape.
Again, though, the lack of fine-scale structure in these visualizations is somewhat surprising, given the quality and diversity of the data.

In this paper, we present a novel approach to studying galaxies that combines elements of several aforementioned techniques but that moves away from the limiting assumptions in their underlying mathematical models. 
Our method, which is an extension of \emph{semi-supervised eigenvectors}~\cite{HM14_JRNL} from \emph{locally-biased machine learning}~\cite{mahoney2012local}, uses ideas from spectral graph theory; and we study the properties of the SDSS Main Galaxy Sample data, using both a traditional global variant as well as a more recently-developed local variant of these methods. 
The global variant reduces to a version of Laplacian eigenmaps~\cite{belkin2003laplacian} and related methods such as diffusion maps~\cite{coifman2006diffusion}, while the local variant exploits recent work on local spectral graph partitioning to engineer locality into the basic global method, while at the same time preserving the strong algorithmic and statistical properties of the global variant~\cite{HM14_JRNL,mahoney2012local}.
Among other things, the method efficiently handles the continuum shape and the spectral lines simultaneously, and the method can be used to explore the data in a qualitatively more refined manner in order to obtain better insights for galaxy studies. 
Since the method is unfamiliar in this application domain, in Section~\ref{sec:methods}, we briefly review the approach; and in Section~\ref{sec:knobs}, we illustrate how several of the key ``knobs'' of the method behave on these data, as a function of data modeling design decisions.
Then, in Section~\ref{sec:global}, we apply the global variant of the method to study the large-structure of the data; and in Section~\ref{sec:local}, we use the local variant of the method to study much finer-scale structure in the data.
Finally, in Section~\ref{sec:conclusion}, we present a brief conclusion.

\section{Global and Local Spectral Embedding Methods}
\label{sec:methods}

The method of locally-biased semi-supervised eigenvectors follows the general approach of recently-popular graph-based machine learning methods: take as input a collection of data points, define some sort of nearest-neighbor graph, and then use eigenvectors and eigenvalues of that graph to define features, parameterize data, perform classification or regression or clustering, etc.
Prior methods that follow this general approach include LLE~\cite{roweis2000nonlinear}, ISOMAP \cite{tenenbaum2000global}, Laplacian eigenmaps~\cite{belkin2003laplacian} and the related diffusion maps~\cite{coifman2006diffusion}, and so on.
These methods are known as nonlinear dimensionality reduction methods since in certain idealized cases they can recover hypothesized nonlinear manifolds and since they provide greater statistical modeling flexibility than straightforward PCA.
It is often more fruitful, however, to view them simply as constructing data-dependent kernels (in the machine learning sense of the word kernel~\cite{SS01-book}) and associated Reproducing Kernel Hilbert Spaces~\cite{HLMS04}.

Among other things, this approach can be used to provide an embedding, i.e., a mapping, of the input data (typically in a very high dimensional Euclidean space) to a very low dimensional Euclidean space.
The embedding is constructed to optimize an objective having to do with connectivity information among data points in an appropriately defined metric. 
For example, Laplacian eigenmaps and diffusion maps use entries of the leading eigenvectors of the Laplacian of the graph or of diffusion processes on the graph to define the embedding.
Under certain assumptions, the Euclidean distance between data points in the embedded space equals a diffusion-based or resistance-based metric on the graph~\cite{belkin2003laplacian,coifman2006diffusion}.
The embedding can then be used for various downstream machine learning and data analysis tasks.
For example, these diffusion-based methods have been explored in the astronomical community, with applications such as photometric~\cite{freeman2009photometric} and spectroscopic~\cite{richards2009exploiting} redshift estimation, and the statistical properties of diffusion embeddings have also been studied previously~\cite{lafon2006diffusion,lee2010spectral}.

\subsection{Constructing Data Graphs}
\label{sec:methods-constructing}

Since our method of locally-biased semi-supervised eigenvectors takes as input a graph, we will start with a description of how data graphs can be constructed from rawer vector-based data, e.g., vectors of astronomical spectra.
Say that we have a collection of $n$ points $\{ x_i \}_{i=1}^n$ in $\R^d$. 
Importantly, since we view the method as constructing a data-dependent kernel, we do \emph{not} assume that the data come from some intrinsically low-dimensional manifold.  
To construct a weighted graph on the data, where the vertices are the data points $x_i$ and the edges represent local connectivity information, we add an edge $(i,j)$ to the graph if $x_j$ is one of $x_i$'s $k$ nearest neighbors or if $x_i$ is one of $x_j$'s $k$ nearest neighbors. 
Note that this ensures the adjacency matrix of the graph is symmetric. 
We then weight each edge with a measure of local similarity given by 
\begin{equation}
	\label{eq:weights}
	w_{ij} = \exp\left( -\frac{\| x_i - x_j \|}{\sigma^2}\right),
\end{equation}
where $\sigma$ is a parameter that controls the amount of ``locality" present in the weight function.
Alternatively, we can add an edge between every pair of points, with weight given by Eqn.~\eqref{eq:weights}, and then sparsify the matrix. 
The $\sigma$ parameter can either be constant across all data points or it can be allowed to vary. 
A useful choice is to set $\sigma^2 = \sigma_i \sigma_j$, where $\sigma_i$ is the distance of point $x_i$ to its $k/2$ nearest neighbor.
Another adaptive choice of $\sigma_i$ was proposed in~\cite{rohrdanz2011determination}. 
Alternatively, one could choose to connect all pairs of points that are closer than some $\epsilon$ distance threshold.
This is the parameterization that we adopt in the remainder of this paper, and we refer to it as ``autotuning" the bandwidth $\sigma$. 
(The graph construction choices surrounding $\epsilon$ or $k$ or $\sigma$ are problem-dependent, and it is known that graph-based machine learning methods are typically quite sensitive to them.  We will discuss the sensitivity of our embeddings to $k$ in Section~\ref{sec:knobs} below.) 

 For the application of interest in this paper, we note that our similarity measure in Eqn.~\eqref{eq:weights} does not take into account the varying signal-to-noise ratio of wavelength bins. 
While we do have access to a per-wavelength uncertainty measure for each data point $x_i$, we choose not to incorporate this information into the weight matrix, and instead we leave this important extension of our method as future work.

\subsection{Locally-biased Semi-supervised Eigenvectors}
\label{sec:methods-sseigs}

Next, we will describe how, given a data graph, we can use global and local spectral methods to construct global and locally-biased embeddings of the nodes of that graph that can then be used for further downstream analysis.
Recall that the usual eigenvectors and eigenvalues of the combinatorial or normalized Laplacian describe successively-orthogonalized directions of maximum variance, and they have strong connections with random walks and diffusion processes on the graph~\cite{chung1997spectral}.
Moreover, they are widely-used in machine learning in general and by nonlinear dimensionality reduction methods in particular.
Here, we will review a methodology to construct so-called \emph{semi-supervised eigenvectors} of a graph Laplacian~\cite{HM14_JRNL}, which is an example of a \emph{locally-biased machine learning} method~\cite{mahoney2012local}.
With appropriate parameter settings, this methodology reduces to the usual global eigenvectors and diffusion-based methods, but with different parameter settings this methodology provides locally-biased analogues of these quantities.

To set notation, let $G=(V,E,w)$ be a connected undirected graph with $n=|V|$ vertices and $m=|E|$ edges, in which edge $(i,j)$ has weight $w_{ij}.$
In the following, $A_G \in \mathbb{R}^{V \times V}$ will denote the adjacency matrix of $G$, while $D_G \in \mathbb{R}^{V \times V}$ will denote the diagonal degree matrix of $G$, i.e., $D_G(i,i)=d_i = \sum_{\{i,j\} \in E} w_{ij}$, the weighted degree of vertex $i$.
The combinatorial Laplacian of $G$ is defined as $L_G \defeq D_G-A_G$, and the normalized Laplacian of $G$ is defined as $\mathcal{L}_G\defeq D_G^{-1/2}L_GD_G^{-1/2}$.
The Laplacian is the symmetric matrix having quadratic form $x^T L_G x = \sum_{ij \in E} w_{ij} (x_i - x_j)^2$, for $x \in  \mathbb{R}^V$.
This implies that $L_G$ is positive semidefinite and that the all-one vector $1 \in  \mathbb{R}^V$ is the eigenvector corresponding to the smallest eigenvalue $0$.
The generalized eigenvalues of $ L_G x = \lambda_i D_G x$ are $0=\lambda_1 < \lambda_2 \leq \cdots \leq \lambda_N$.
We will use $\nu_2$ to denote smallest non-trivial eigenvector, i.e., the eigenvector corresponding to $\lambda_2$; $\nu_3$ to denote the next eigenvector; and so on.
We will overload notation to use $\lambda_2(A)$ to denote the smallest non-zero generalized eigenvalue of $A$ with respect to $D_G$. 

Given this notation, consider the left panel of Figure~\ref{fig:objective}.
The leading nontrivial global eigenvector $\nu_2$ of the normalized Laplacian $\mathcal{L}_G$ (or, equivalently, the leading nontrivial generalized eigenvectors of $L_G$) is the solution to the problem \textsc{GlobalSpectral}.  
The next eigenvector $\nu_3$ is the solution to \textsc{GlobalSpectral}, augmented with the constraint that $x^TD_G\nu_2=0$; and in general the $t^{th}$ generalized eigenvector of $L_G$ is the solution to \textsc{GlobalSpectral}, augmented with the constraints that $x^TD_G\nu_i=0$, for $i\in\{2,\ldots,t-1\}$.
It is these vectors that are widely-used in machine learning and data analysis, e.g., they are used to construct embeddings in Laplacian eigenmaps, and it is regularized variants of these vectors (computed from random walks) that are used to construct embeddings in diffusion maps.
In particular, these eigenvectors can be used to define a low-dimensional embedding~via
\begin{equation}
	\label{eq:define-global-embed}
	x_i \mapsto (\nu_2^i, \nu_3^i, \ldots, \nu_k^i),
\end{equation}
where the embedding dimension $k$ is a parameter to be chosen and the notation $\nu_j^i$ represents the $i^\textup{th}$ element of the vector $\nu_j$.
(Since standard results from spectral graph theory show that the constant vector of all ones is an eigenvector with eigenvalue one~\cite{chung1997spectral}, the first eigenvector $\nu_1$ is not used in the embedding.) 
In the embedding space, the standard Euclidean distance between two points is proportional to the average length of a random walk starting at one point and reaching the second. 
In this sense, the embedding given by these eigenvectors preserves ``connectivity'' information about the original data. 
For further details on the theory of these methods, we refer the reader to~\cite{belkin2003laplacian,coifman2006diffusion}.

\begin{figure*}[t]
\begin{minipage}[t]{0.3\linewidth}
\centering
\textsc{GlobalSpectral}\\
\;
\begin{align*}
\text{min}_{x\in\mathbb{R}^{n}} \quad & x^T L_G x \\
\text{s.t}\quad & x^T D_G x = 1 \\
& x^T D_G 1 = 0 
\end{align*}
\end{minipage}
\begin{minipage}[t]{0.3\linewidth}
\centering
\textsc{LocalSpectral}\\
\;
\begin{align*}
\text{min}_{x\in\mathbb{R}^{n}} \quad & x^T L_G x \\
\text{s.t}\quad & x^T D_G x = 1 \\
& x^T D_G 1 = 0 \\
& x^T D_G s \geq \sqrt{\kappa}
\end{align*}
\end{minipage}
\hspace{0.5cm}
\begin{minipage}[t]{0.3\linewidth}
\centering
\textsc{Generalized LocalSpectral}
\begin{align*}
\text{min}_{x\in\mathbb{R}^{n}} \quad & x^T L_G x \\
\text{s.t}\quad & x^T D_G x = 1 \\
& x^T D_G X = 0 \\
& x^T D_G s \geq \sqrt{\kappa}
\end{align*}
\end{minipage}
\caption{%
Left: The usual \textsc{GlobalSpectral} partitioning problem; the vector achieving the optimal solution is $\nu_2$, the leading nontrivial 
generalized eigenvector of $L_G$ with respect to $D_G$.
Middle: The \textsc{LocalSpectral} problem, which includes a locality constraint, was originally introduced in~\cite{mahoney2012local}. 
Right: The \textsc{Generalized LocalSpectral} problem, which includes both the locality constraint and a more general orthogonality constraint, was originally introduced in~\cite{HM14_JRNL}.
The main algorithm for computing locally-biased semi-supervised eigenvectors will iteratively compute the solution to \textsc{Generalized LocalSpectral} for a 
sequence of $X$ constraint matrices~\cite{HM14_JRNL}.
}
\label{fig:objective}
\end{figure*}

While these embeddings provide an intuitive low-dimensional representation of high-dimensional data, they suffer from their fundamentally global nature. 
By this we mean that the embeddings are given by the eigenvectors of a suitably defined matrix, which can themselves be viewed as the solution of the global optimization problem \textsc{GlobalSpectral}, and as a consequence of the global orthogonality requirements, local structure and local heterogeneities tend to be lost in such an embedding. 
One (weak) way to tune this homogenizing effect is via the nearest-neighbor parameter $k$; 
another (stronger) way to recover local information is to enforce an additional constraint in the optimization program.

Motivated by this observation, we will define locally-biased analogues of these quantities in two steps: first, we will define the locally-biased analogue of the leading nontrivial eigenvector; and second, we will define the locally-biased analogue of the remainder of the eigenvectors.
(The reason for this two-step process is that the leading nontrivial eigenvector and the remaining eignevectors have slightly different interpretations with respect to random walks; describing this is beyond the scope of this paper, but details can be found in~\cite{mahoney2012local,HM14_JRNL}.)
To do so, assume that we are given as input a (possibly weighted) data graph $G=(V,E,w)$, an indicator vector $s$ of a small ``seed set'' of nodes, a correlation parameter $\kappa \in [0,1]$, and a positive integer $k$.
(This $k$ will be the number of eigenvectors to be chosen, and thus it is different than the $k$ used to define the number of nearest neighbors.)
Then, informally, we would like to construct $k$ vectors that satisfy the following bicriteria:
first, each of these $k$ vectors is well-correlated with the input seed set; and 
second, those $k$ vectors describe successively-orthogonalized directions of maximum variance, in a manner analogous to the leading $k$ nontrivial global eigenvectors of the graph Laplacian.
We emphasize that---in the same way as $G$ and $k$ are input to the usual global spectral methods---here the seed set $s$ of nodes, the integer $k$, and the correlation parameter $\kappa$ are also part of the input.  
They should be thought of as being available in a semi-supervised manner.

Given this, consider the middle panel of Figure~\ref{fig:objective}.
This presents \textsc{LocalSpectral}, an optimization problem that was introduced in~\cite{mahoney2012local} which extends \textsc{GlobalSpectral} by including a constraint that the solution be well-correlated with an input seed set.
We can assume (without loss of generality) that $s$ is properly normalized and orthogonalized so that $s^T D_{G} s =1$ and $s^T D_{G} 1 =0$.
Also, while $s$ can be a general unit vector orthogonal to $1$, it may be helpful to think of $s$ as the indicator vector of one or more vertices in $V$.
The solution to \textsc{LocalSpectral} may be interpreted as a locally-biased version of the second eigenvector of the Laplacian.
Importantly, it's solution can be computed efficiently as the solution to a set of linear equations that generalize the popular diffusion-based Personalized PageRank procedure (and thus it can also be computed via a random walk process).
In~particular,
\begin{equation}
x_1^{*} = c \left( L_G - \gamma_1 D_G \right)^{+} D_Gs   ,
\label{eqn:first-step}
\end{equation}
where $\gamma_1$ is chosen to saturate the correlation constraint.
By performing a sweep cut and appealing to a variant of Cheeger's inequality, this locally-biased eigenvector can be used to perform locally-biased spectral graph partitioning~\cite{mahoney2012local}.
Observe that, for $\kappa=0$, this coincides with the usual \textsc{GlobalSpectral} objective, while for $\kappa > 0$, this produces solutions that are biased toward the seed vector $s$.

Finally, consider the right panel of Figure~\ref{fig:objective}.
For this \textsc{Generalized LocalSpectral} problem, in addition to the previous setup, we are also given an $n \times \nu$ constraint matrix $X$ that may be assumed to be an orthogonal matrix.
In words, this problem asks us to find a vector $x \in \mathbb{R}^{n}$ that minimizes the variance $x^TL_Gx$ subject to several constraints: that $x$ is unit length; that $x$ is orthogonal to the span of $X$; and that $x$ is $\sqrt{\kappa}$-well-correlated with the input seed set vector $s$.
To compute these locally-biased semi-supervised eigenvectors, we follow~\cite{HM14_JRNL} and iteratively compute the solution to \textsc{Generalized LocalSpectral}, updating $X$ to contain the already-computed semi-supervised eigenvectors.
In particular, to compute the first semi-supervised eigenvector, we let $X=1$ (i.e., the $n$-dimensional all-ones vector, which is the trivial eigenvector $L_G$, in which case $X$ is an $n \times 1$ matrix), in which case the first nontrivial semi-supervised eigenvector is given by Eqn.~(\ref{eqn:first-step}).
To compute each subsequent semi-supervised eigenvector, we let the columns of $X$ consist of $1$ and the other semi-supervised eigenvectors found in each of the previous iterations.
See~\cite{HM14_JRNL} for details on how  \textsc{Generalized LocalSpectral} can be solved efficiently in terms of a sequence of  linear equations or constrained eigenvalue problems.
We simply note that, in practice, the correct value of $\gamma$ is unknown, and one performs a binary search over the range $(-\infty,\lambda_2(G))$ until the correlation constraint is saturated.
Each stage of this binary search involves solving a system of linear equations, which can be done efficiently using iterative methods such as conjugate gradient. 

We should note that, in this paper, we are interested in understanding local versus global properties of the data, and so we adopt an ``exploratory'' approach.
For other downstream learning tasks, e.g., classification or regression, various model-selection methods can be used to select $k$, $\sigma$, $\gamma$, $\kappa$, etc.~\cite{friedman2001elements}.
Extending the methodology of locally-biased semi-supervised eigenvectors to model selection and other related statistical questions, e.g., those considered in~\cite{richards2009exploiting,vanderplas2009reducing}, is straightforward.

\subsection{Regularization and Diffusion Interpretation}

While the previous discussion has been in terms of constrained eigenvectors of the Laplacian, one can also describe these global and local spectral methods in terms of random walks, and this often aids intuition and computation.
Consider, first, the usual global method.
We can construct a matrix $M$ related to the transition matrix of a ``lazy" random walk on the data as follows:
\begin{equation}
        \label{eq:define-M}
        M \defeq \frac{1}{2} D^{-1/2}\left( D+W \right)D^{-1/2}.
\end{equation}
If both factors of $D^{-1/2}$ acted on the right of $D+W$, this would correspond to the transition matrix $T$ of a random walk on the data, where at each step the walker flips a fair coin: heads, he stays put; tails, he jumps to a random neighbor of the current point $i$ with probabilities proportional to the $i^\textup{th}$ row of $W$.
Note that this transition matrix $T$ differs by a similarity transformation from $M$ as defined in Eqn.~\eqref{eq:define-M}, given by $T = D^{1/2} M D^{-1/2}$.
The two matrices thus share eigenvalues, and their eigenvectors differ by multiplication with a diagonal matrix ($D^{\pm 1/2}$).
Moreover, this eigensystem of $M$ is directly related to the eigensystem of the graph Laplacian $L$~\cite{chung1997spectral}.
The normalization we choose ensures that $M$ is symmetric, while the offset by $D$ (the ``laziness" of the walk) ensures that $M$ is positive semidefinite.
These properties are desirable for the numerical computation of the leading eigenvalues and corresponding eigenvectors using iterative methods.

We should also note that there is a natural ``regularization'' interpretation underlying our construction of our locally-biased semi-supervised eigenvectors and that this is intimately related to working with locally-biased random walks on the data graph $G$.
To see this, recall that the first step of our algorithm can be computed as the solution of a set of linear equations of the form of Eqn.~(\ref{eqn:first-step}), for some normalization constant $c$ and some $\gamma$.
The quantity $\left( L_G - \gamma D_G \right)^{+}$ can be interpreted as a ``regularized'' version of the pseudoinverse of $L$, where $\gamma\in(-\infty,\lambda_2(G))$ serves as the regularization parameter.
This regularization interpretation has recently been made precise~\cite{MO11-implementing,PM11}.
Alternatively, the quantity quantity $\left( L_G - \gamma D_G \right)^{+}$ is exactly a generalization of the usual PageRank vector that involves ``random walkers'' who uniformly (or non-uniformly, in the case of Personalized PageRank) ``teleport'' with a probability $\alpha\in(0,1)$.
As described in~\cite{mahoney2012local},  choosing $\alpha\in(0,1)$ corresponds precisely to choosing $\gamma \in (-\infty,0)$.
In this case, the random walk interpretation is that one performs random walks consisting of a small number of steps, starting from a locally-biased seed node (as opposed to a large number of steps starting from an arbitrary seed node, which yields the usual global diffusion-based methods).
For readers more comfortable with Laplacian eigenmaps or diffusion maps (or other diffusion-based machine learning methods), we simply note that the locally-biased semi-supervised eigenvectors framework can be used to provide locally-biased variants of these methods; we refer the interested reader to~\cite{mahoney2012local,HM14_JRNL} for details. 

Given these connections, our computations of global diffusion embeddings were performed in MATLAB using a modified version of the \texttt{DiffusionGeometry} package of Bremer and Maggioni~\cite{coifman2005geometric,bremer2006diffusion}; and our computations of local embeddings were performed in MATLAB using the \texttt{sseigs\_demo} package of Hansen and Mahoney~\cite{hansen2012sseigs,HM14_JRNL}.

\section{Initial Illustration of the Method on SDSS Data}
\label{sec:knobs}

In this section, we describe the data we will consider as well as some of the most important ``knobs'' of the global and local spectral embedding methods described in Section~\ref{sec:methods}.
Details such as these are often relegated to an appendix or a methods section, but we place them here since the behavior of the method depends on these knobs in important ways.

\subsection{The Main Galaxy Sample}
\label{subsec:data-preprocessing}

The Main Galaxy Sample (MGS) has become a testbed for a wide range of astronomical studies.
There are several reasons for this: due to the well-understood selection function, due to the large volume of high-quality data, and due to the prior systematic analyses that serve as a reference for new techniques.
The definition of the MGS~\cite{strauss2002mgs} uses a single brightness limit of 17.77 Petrosian magnitudes. 
In the local (and recent) universe, this translates to a complete sample of galaxies that is dominated by red ellipticals. 
In our study, we use the entire rest frame wavelength range between 3450 {\AA} and 8350 {\AA}.
Starting from the spectrophotometrically calibrated spectra, our only preprocessing consists of dealing with gaps in the wavelength coverage. 
The missing parts of the spectra are simply filled-in based on the best fit linear combination using eigenspectra from a prior PCA analysis following~\cite{yip04a}. 
This simplifies the implementations of our method, but we note that our approach is also capable of dealing with incomplete data, e.g., similarly to gappy PCA~\cite{connolly95b}.

\subsection{Effect of Using $\eps$-NN or $k$-NN Parameterization}
\label{subsec:density}

An important aspect of our method involves the choice of nearest neighbors (NNs), and so we pause here to remark on the extremely heterogenous density of the sample, and how it affects our choice of embedding parameters. 

In many applications, it is of interest, e.g., for reasons of interpretability, to use a constant bandwidth $\sigma$ over all data points, i.e., choose NN based on an $\eps$-NN threshold. 
A naive implementation of this would be computationally infeasible, even for moderately-large data, since the majority of galaxies lie in a very dense region of wavelength space. 
The matrices whose eigenvectors define our embeddings become very dense if we require galaxies within a fixed distance to be connected; and, indeed, such matrices would not fit in memory on the machines used. 

\begin{figure}
	\centering
	\includegraphics[width=.4\textwidth]{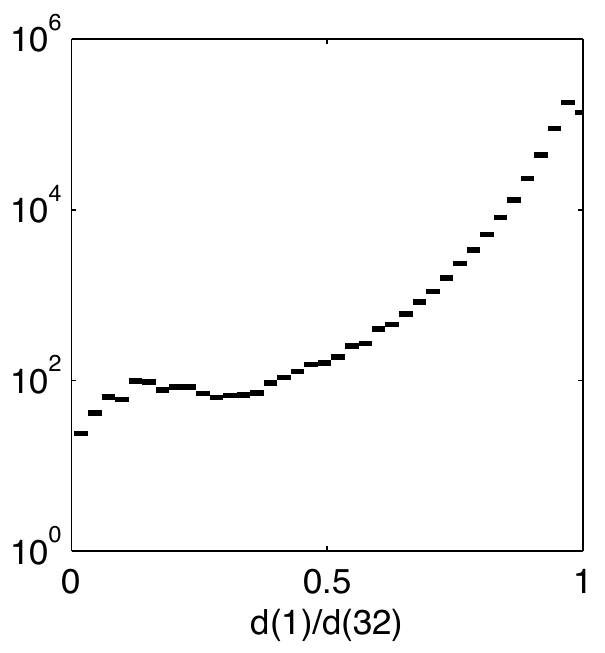}
	\caption{Histograms of ratios of distances to first to 32nd nearest neighbors, over the full 517K spectra data set. The vast majority of galaxies lie in very dense regions of the parameter space (corresponding to ratios near unity), while a non-negligible fraction lie in very sparsely populated regions (the bumps near zero).}
	\label{fig:density-histos}
\end{figure}

One view of this heterogeneity is presented in Figure~\ref{fig:density-histos}, which shows a histogram (in logarithmic scale) of the ratios of first to 32nd NNs over the entire data set. It is clear from the figure that the vast majority of galaxies lie in very dense regions of space (for which this ratio is close to one), while a non-negligible fraction lie in more sparsely populated regions (for which the ratio is less than one half.)

In this paper, we choose to autotune the bandwidths and fix the number of NN edges, in order to keep our computations tractable on the hardware at our disposal. 
We note that~\cite{coifman2006diffusion} proposed an alternative normalization of the diffusion operator of Eqn.~\eqref{eq:define-M} (which they term the ``Beltrami'' normalization) that removes the effect of the sampling density on the embeddings. 
Our computations confirmed the amelioration of this effect to some degree, but at the cost of increased runtime due to the slower decay of the spectrum associated with the Beltrami normalization. 
For this reason we keep the former normalization of Eqn.~\eqref{eq:define-M}, termed ``Markov'' normalization, for the remainder of this paper.

\subsection{Effect of Global via Nearest-neighbors}
\label{subsec:effects-parameters}

Next, we discuss the choice of $k$, the number of NN edges when using the  $k$-NN rule, on the constructed graphs and their embeddings. 
While the effects of the choice of $k$ can be highly problem-dependent, and while the choice of $k$ ultimately should be determined by a downstream astronomical model selection criteria (e.g., optimizing mean-squared error, or a precision-recall metric if one is performing classification, or obtaining qualitative insight into rare galaxy types of interest if one is interested in a more targeted goal), we have noticed several general trends of interest. 
First, if one chooses $k$ to be larger, then one tends to identify well the large-scale or global structure in the data, while washing out or homogenizing small-scale or local structure.
Second, and relatedly, if one chooses $k$ to be smaller, then local or small-scale structure tends to be highlighted, but since the constructed graphs are much sparser and noisier, it is more difficult to draw sufficient statistical strength to identify as reliably global or large-scale structure.

\begin{figure}
      \begin{center}
      \subfigure[]{
         \includegraphics[width=0.45\textwidth]{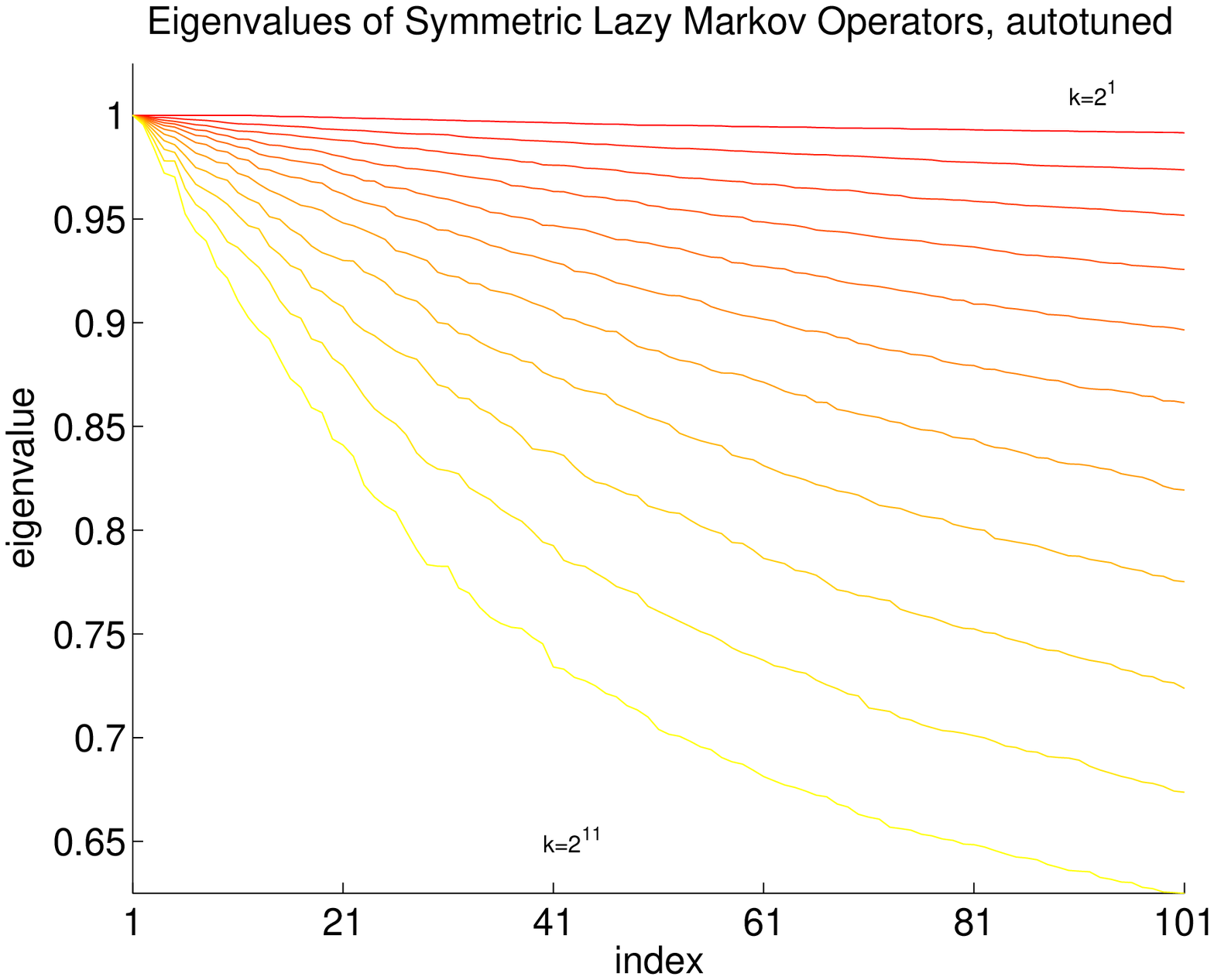}
	 \label{fig:eigenvalues-A}
      } \qquad 
      \subfigure[]{
         \includegraphics[width=0.45\textwidth]{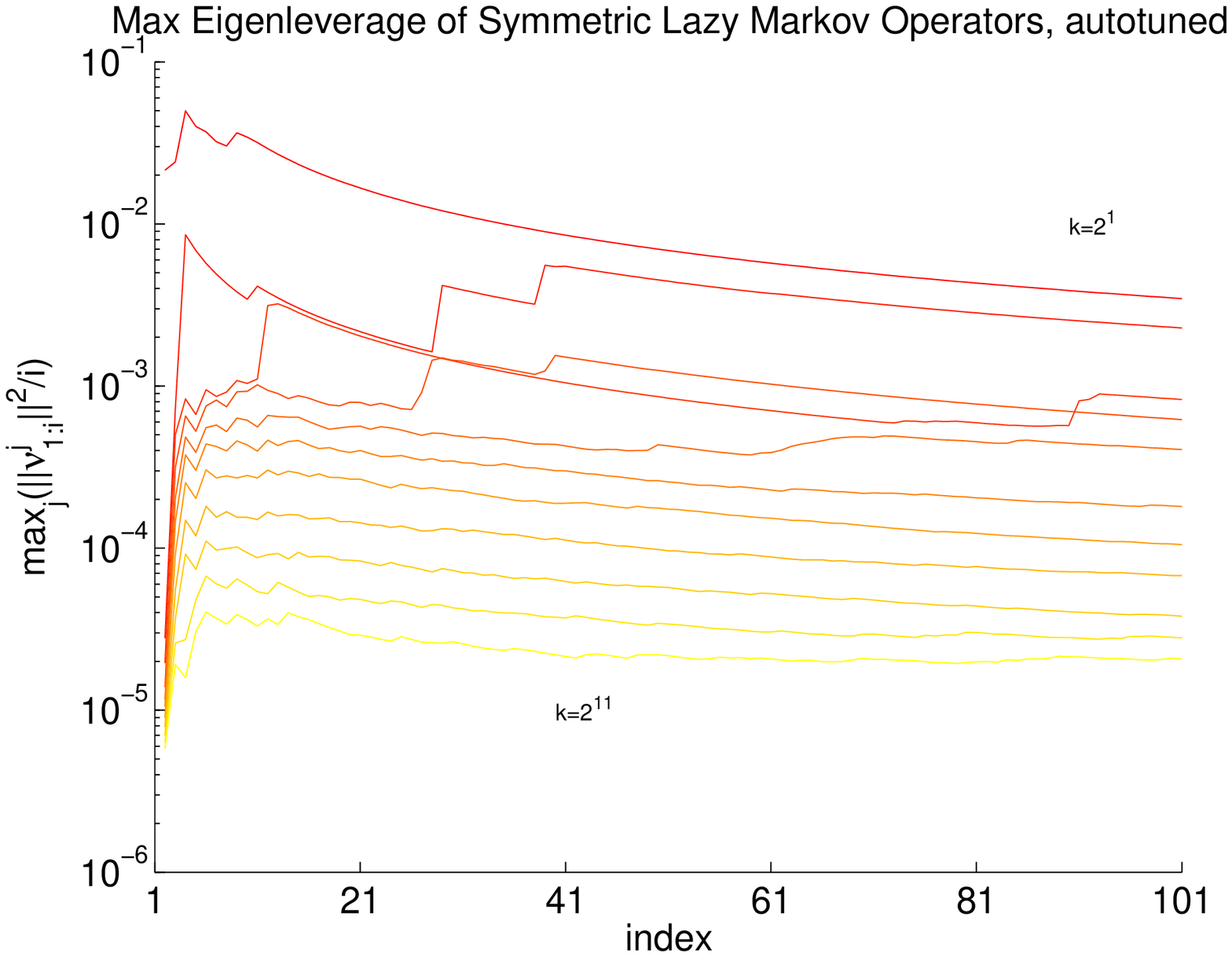}
	 \label{fig:eigenvalues-B}
      } 
      \end{center}
	\caption{(a) Top 101 eigenvalues of the lazy Markov operator with autotuned bandwidths, for $k=2^1$ (red) to $k=2^{11}$ (yellow). (b) Max-to-median ratio of eigenvector norms, as a function of embedding dimension for the lazy Markov operator with autotuned bandwidths, for $k=2^1$ (red) to $k=2^{11}$ (yellow).}
	\label{fig:eigenvalues}
\end{figure}

To make these informal observations somewhat more quantitative, we describe next the uniformity properties of both the eigenvalues as well as the eigenvectors used in our embeddings.
In Figure~\ref{fig:eigenvalues-A}, we plot the decay of the top 101 eigenvalues of the lazy Markov operator with autotuned bandwidths, for values of $k$ ranging from $2^1$ to $2^{11}$ by powers of two. 
From the figure, we can see that as $k$ is increased and thus as more edges are added to the graph, the rate at which the eigenvalues decay increases, i.e., the matrix is more well-embeddable in a low-dimensional space. 
Next, in Figure~\ref{fig:eigenvalues-B}, we plot the ratio of the largest eigenvector norm to the median eigenvector norm, as a function of the index of the eigenvectors. 
Specifically, if $\nu_{1:i}^j$ is the embedding of spectrum $j$ on eigenvectors 1 to $i$, we plot the quantity
\begin{equation}
\label{eq:max-median}
	\frac{1}{i}\max_{1\le j \le N} \left( \left\| \nu_{1:i}^j \right\| \right)
\end{equation}
as a function of $i$. 
This quantity is a measure of the non-uniformity of the distribution of mass on the eigenvectors, in a manner similar to leverage scores in linear dimension reduction~\cite{yip2014objective,chatterjee2009sensitivity,CUR_PNAS}. 
In general, the eigenvectors become more uniform with increasing $k$, and local heterogeneities---as captured by localized eigenvectors---become less prominent with the inclusion of additional edges. 
In Table~\ref{tab:effect-of-k}, we give the values of several measures of matrix complexity as $k$ is changed. 
We note in particular that as the number of nearest-neighbor edges increases, both the stable rank and the leverage scores decrease monotonically. 
This accords with the intuition mentioned previously that higher connectivity tends to ``smooth out'' the embeddings. 

It is perhaps interesting to note that for these values of $k$, the Markov matrix is not particularly well-approximated by a low-rank matrix. 
Nevertheless, the leading eigenvectors of $M$ do correlate well with physical intuition, and they do provide meaningful low-dimensional representations of the data.
We should also note that these effects of increasingly-localized eigenvectors and slower eigenvalue decay as NN data graphs are made sparser has been observed previously in connection with the Nystr\"{o}m method in large-scale machine learning applications; see the statistics on the sparsified radial basis function kernels in the analysis of~\cite{GM15_NYSTROM_JRNL}.
Our results are consistent with this prior work~\cite{GM15_NYSTROM_JRNL} that demonstrated that sparser graphs (within a parameterized family such as with radial basis function kernels or $k$-NN graphs) are much less ``nice'' in the sense that their eigenvalues decay more slowly and that their eigenvectors are more localized.

\begin{table}
	\centering
	\begin{tabular}{l|l|l|l|l|l|l|l}
	$k$ & \% nnz & $\left\lceil \frac{\| M \|_\textup{F}^2}{\| M \|_2^2} \right\rceil$ &
	$\frac{\lambda_{101}}{\lambda_{100}}$ & $100\frac{\| M-M_{100} \|_2}{\| M \|_2}$ & 
	$100\frac{\| M-M_{100}\|_\textup{F}}{\| M \|_\textup{F}}$ & 
	$100\frac{\| M-M_{100}\|_*}{\| M \|_*}$ & 
	\parbox{.85in}{{\footnotesize 100th largest leverage score}} \\ \hline 
	2 & 0.000963 & 317866 & 1.0000 & 99.1685 & 99.9844 & 99.9754 &    604 \\
    4 & 0.001730 & 252304 & 0.9999 & 97.4220 & 99.9807 & 99.9726 &    347 \\
    8 & 0.003263 & 201239 & 0.9996 & 95.2012 & 99.9765 & 99.9698 &    182 \\
    16 & 0.006323 & 168355 & 0.9994 & 92.7084 & 99.9727 & 99.9674 &    138 \\
    32 & 0.012431 & 149618 & 0.9983 & 89.6513 & 99.9704 & 99.9661 &     92 \\
    64 & 0.024609 & 139635 & 0.9995 & 86.1339 & 99.9696 & 99.9657 &     65 \\
    128 & 0.048867 & 134492 & 0.9995 & 81.8590 & 99.9701 & 99.9661 &     44 \\
    256 & 0.097110 & 131889 & 0.9995 & 77.3375 & 99.9716 & 99.9669 &     30 \\
    512 & 0.192818 & 130582 & 0.9966 & 72.1517 & 99.9736 & 99.9682 &     18 \\
    1024 & 0.382121 & 129932 & 0.9990 & 67.0726 & 99.9759 & 99.9697 &     12 \\
    2048 & 0.755086 & 129610 & 0.9997 & 62.2042 & 99.9785 & 99.9714 &      9
   	\end{tabular}
	\caption{Several measures of matrix complexity as the nearest-neighbor parameter $k$ is varied.  As $k$ is decreased, the graphs are sparser, and both eigenvector-based and eigenvalue-based ``niceness'' metrics for the matrices are much worse (in agreement with the results of~\cite{GM15_NYSTROM_JRNL}).}
	\label{tab:effect-of-k}
\end{table}

\begin{figure}
      \begin{center}
      \subfigure[]{
         \includegraphics[width=.30\textwidth]{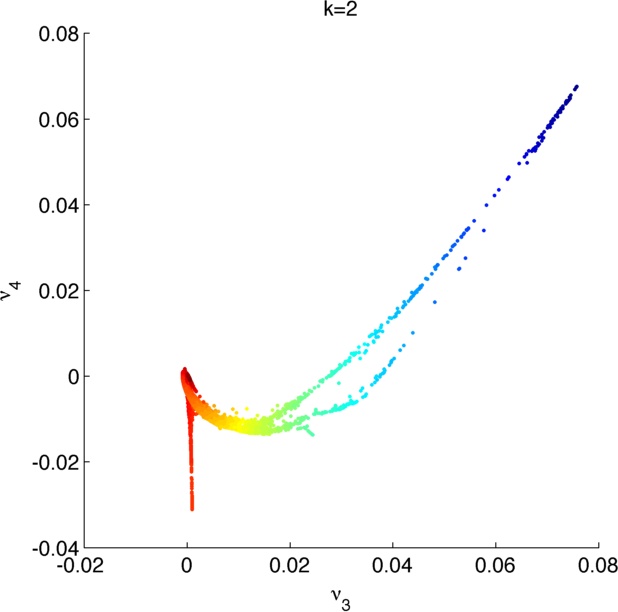}
	 \label{fig:effect-of-k-markov-A}
      } 
      \subfigure[]{
         \includegraphics[width=.30\textwidth]{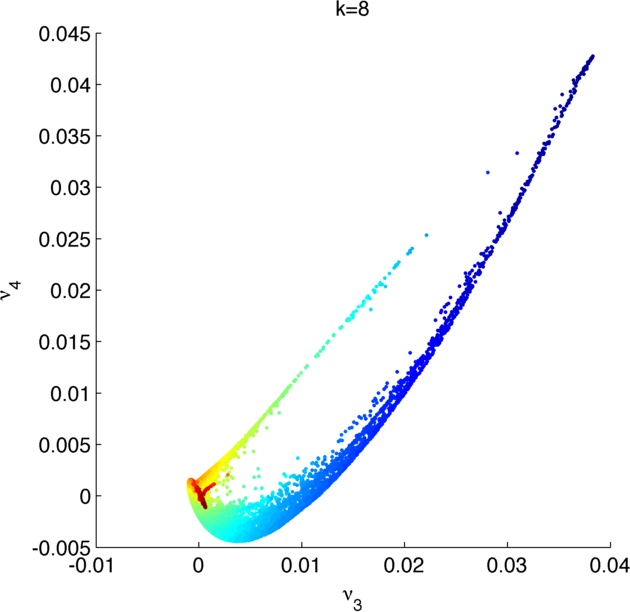}
	 \label{fig:effect-of-k-markov-B}
      } 
      \subfigure[]{
         \includegraphics[width=.30\textwidth]{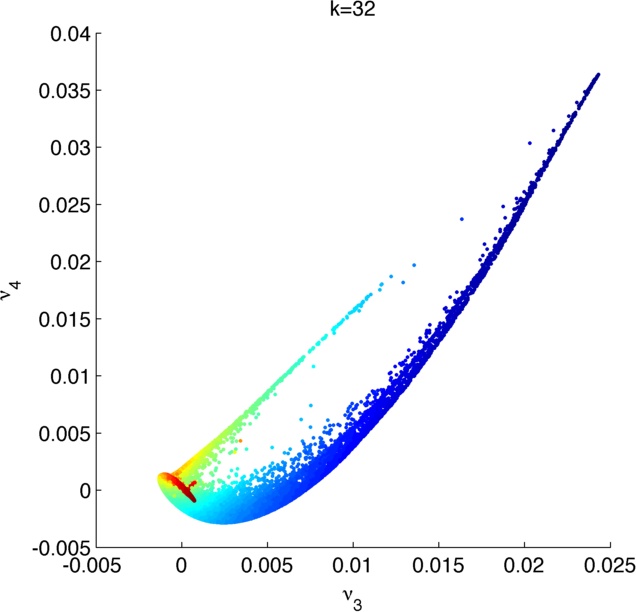}
	 \label{fig:effect-of-k-markov-C}
      } \\
      \subfigure[]{
         \includegraphics[width=.30\textwidth]{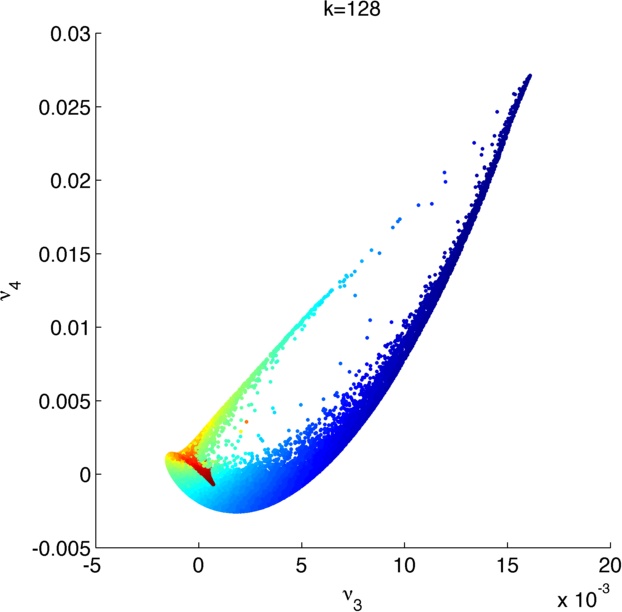}
	 \label{fig:effect-of-k-markov-D}
      } 
      \subfigure[]{
         \includegraphics[width=.30\textwidth]{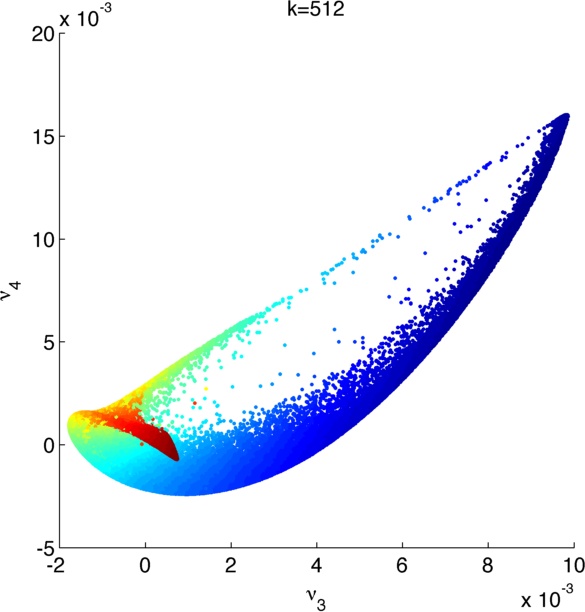}
	 \label{fig:effect-of-k-markov-E}
      } 
      \subfigure[]{
         \includegraphics[width=.30\textwidth]{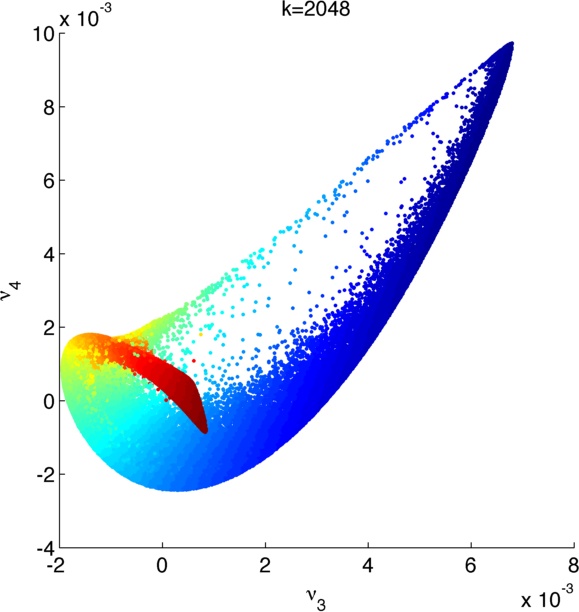}
	 \label{fig:effect-of-k-markov-F}
      } 
      \end{center}
\caption{Embedding on third and fourth eigenvectors of the lazy Markov operator with autotuned bandwidths, for $k=2^n, \; n \in \{ 1,3,5,7,9,11\}$. Points are color-coded by the value of the second eigenvector.}
\label{fig:effect-of-k-markov}
\end{figure}

Finally, in Figure~\ref{fig:effect-of-k-markov}, we show the embeddings of the full data set on the third and fourth eigenvectors (we will consider other eigenvectors below) of the lazy Markov operator with autotuned bandwidths, for $k$ ranging from $2^1$ to $2^{11}$ by factors of four. 
The points are color-coded by the value of the second eigenvector. 
These plots hide density information (one example of which we will present below), but they illustrate that as $k$ increases, the data points become visually more compressed, with more linkages between different features of the point cloud. 
In particular, the red part of the embedding varies considerably as $k$ is adjusted, and small-scale local structure such as the cyan ``heel'' pointing downward and to the right for $k=2$ is lost for larger values of $k$.
In Section~\ref{sec:global}, we will argue that physical considerations imply that the connection between the dark blue and cyan points is an artifact of the method, and that the value $k=32$ seems to provide a more meaningful embedding; and in Section~\ref{sec:local}, we will describe in more detail how to obtain insight about the small-scale or local properties of this. 

\section{Global Structure via Global Embeddings}
\label{sec:global}

In this section, we provide examples of how our method can be used as an exploratory tool to identify large-scale global structure in the SDSS data.
This should provide some intuition regarding the physical interpretation of the leading non-trivial eigenvectors of the Markov operator $M$, and it will illustrate the strength of our approach at highlighting properties of the data. 
In particular, in Section~\ref{subsec:density-embeddings}, we illustrate the density of galaxies in the embedding from several different angles;
in Section~\ref{subsec:embedding-2vs4}, we provide an illustration of the data that highlights the continuum shape and line strength;
in Section~\ref{subsec:embedding-2vs5}, we provide an illustration of the data that discriminates particularly well red elliptical galaxies;
in Section~\ref{subsec:pca-eigs}, we illustrate the data embedded on the PCA mixing angles; 
in Section~\ref{subsec:bpt}, we compare our method with the insight that can be obtained with the BPT diagrams;
in Section~\ref{subsec:embedding-class}, we compare the results of our method with the SDSS classification; and finally 
in Section~\ref{subsec:redshift}, we discuss a peculiar aspect of one of the figures and explain it in terms of an artifact of the experimental measurements.

\subsection{Density of Galaxies}
\label{subsec:density-embeddings}

\begin{figure}
      \begin{center}
      \subfigure[]{
         \includegraphics[width=0.30\textwidth]{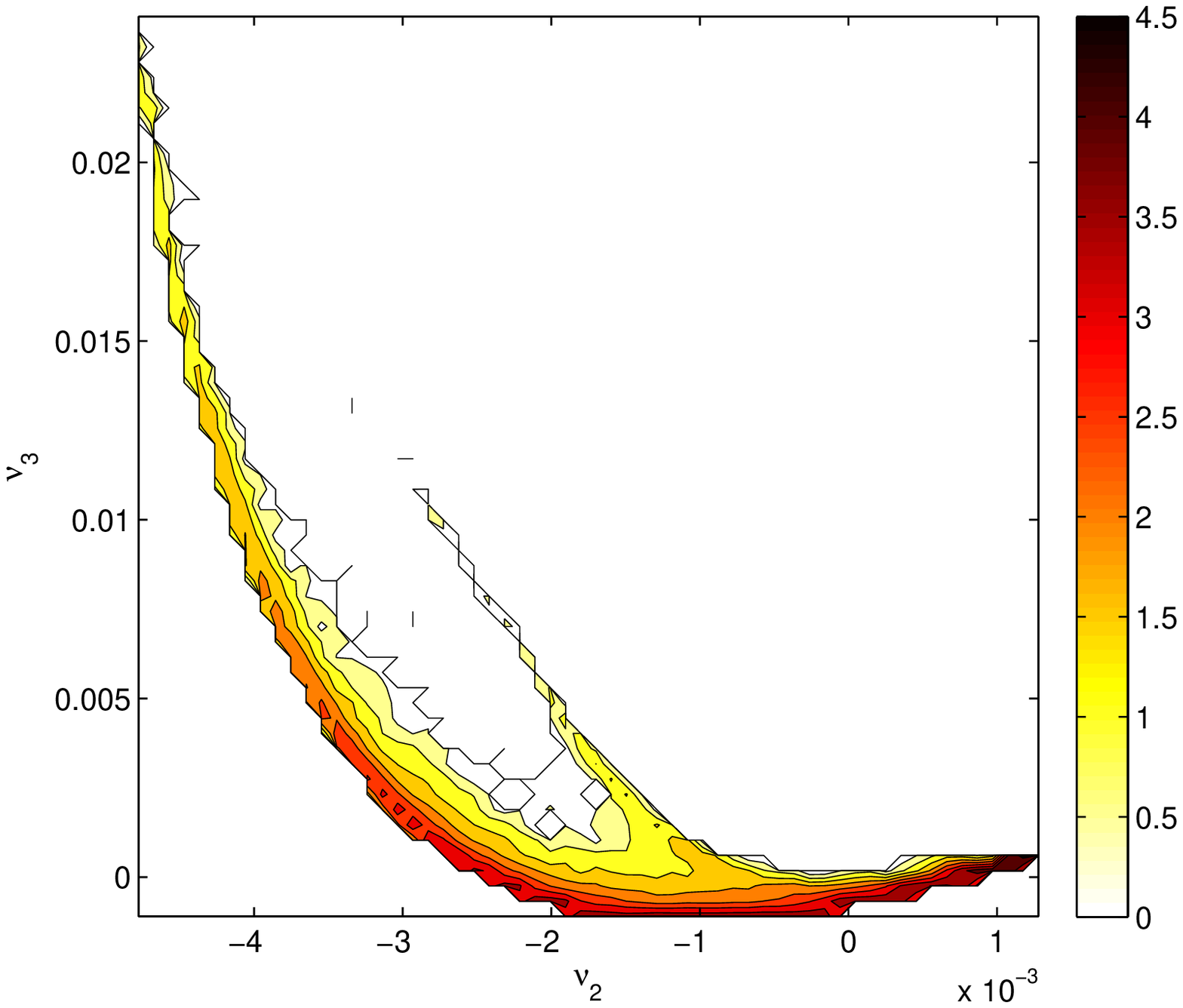}
	 \label{fig:markov-densities-A}
      } 
      \subfigure[]{
         \includegraphics[width=0.30\textwidth]{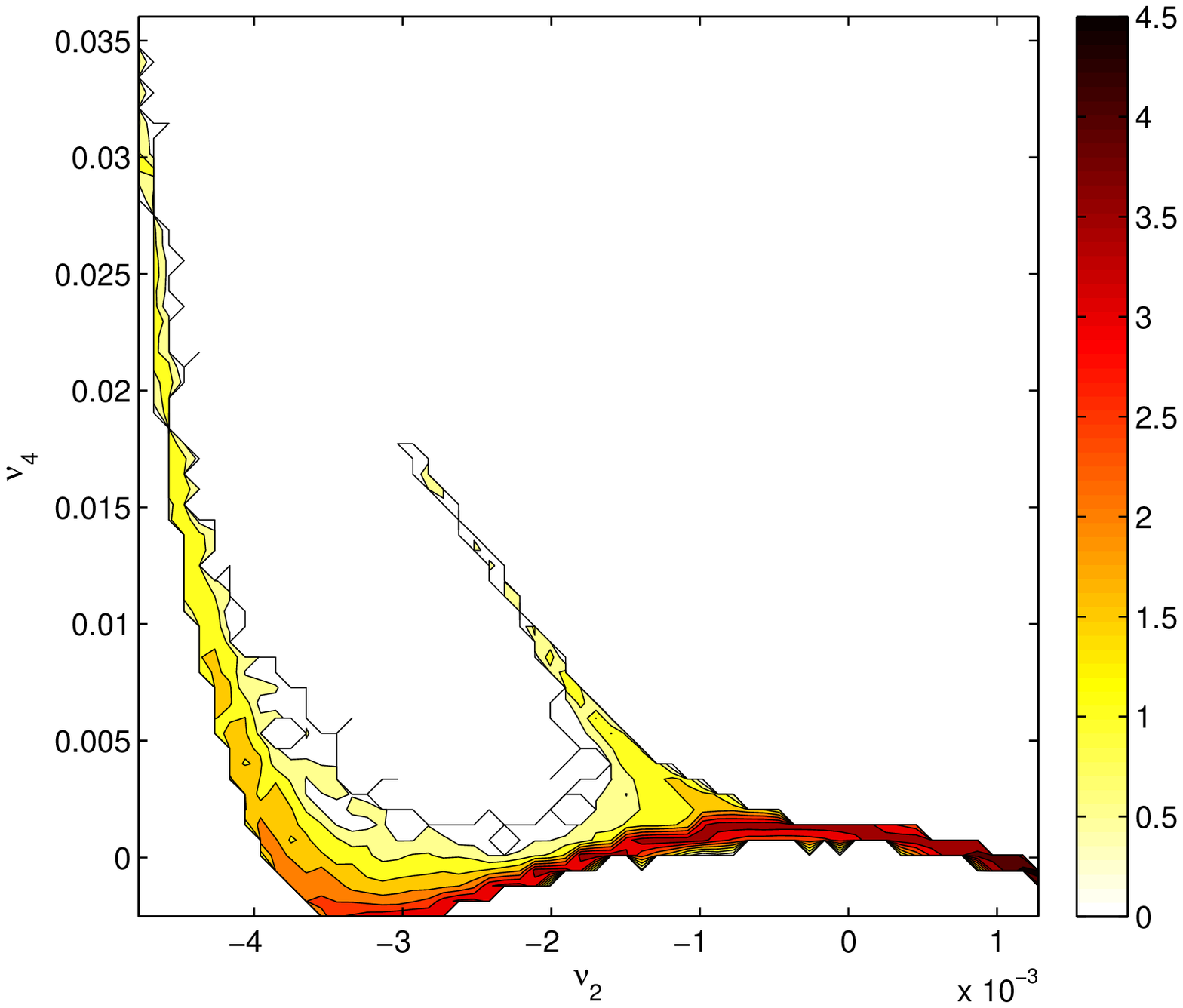}
	 \label{fig:markov-densities-B}
      } 
      \subfigure[]{
         \includegraphics[width=0.30\textwidth]{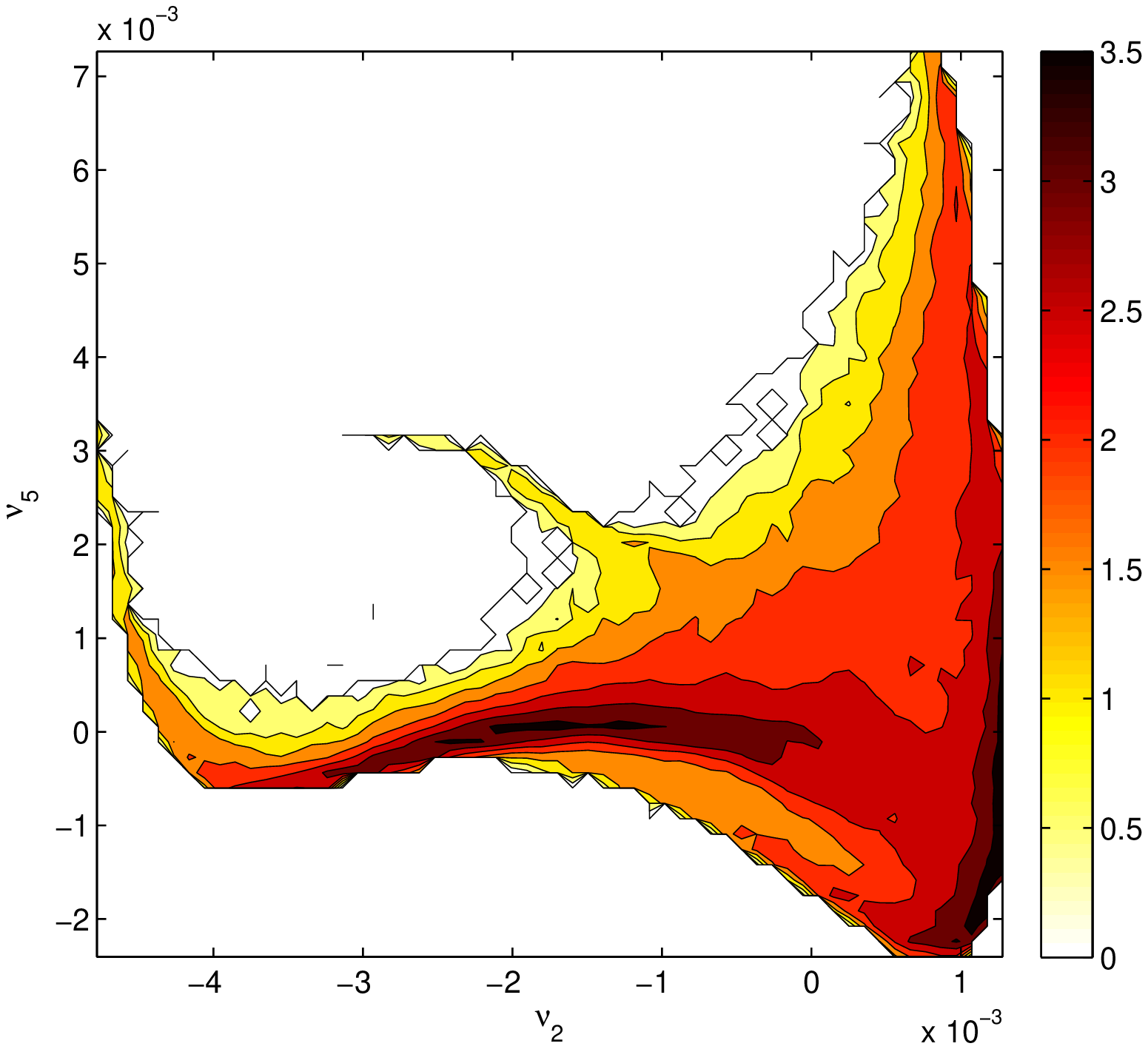}
	 \label{fig:markov-densities-C}
      } \\
      \subfigure[]{
         \includegraphics[width=0.30\textwidth]{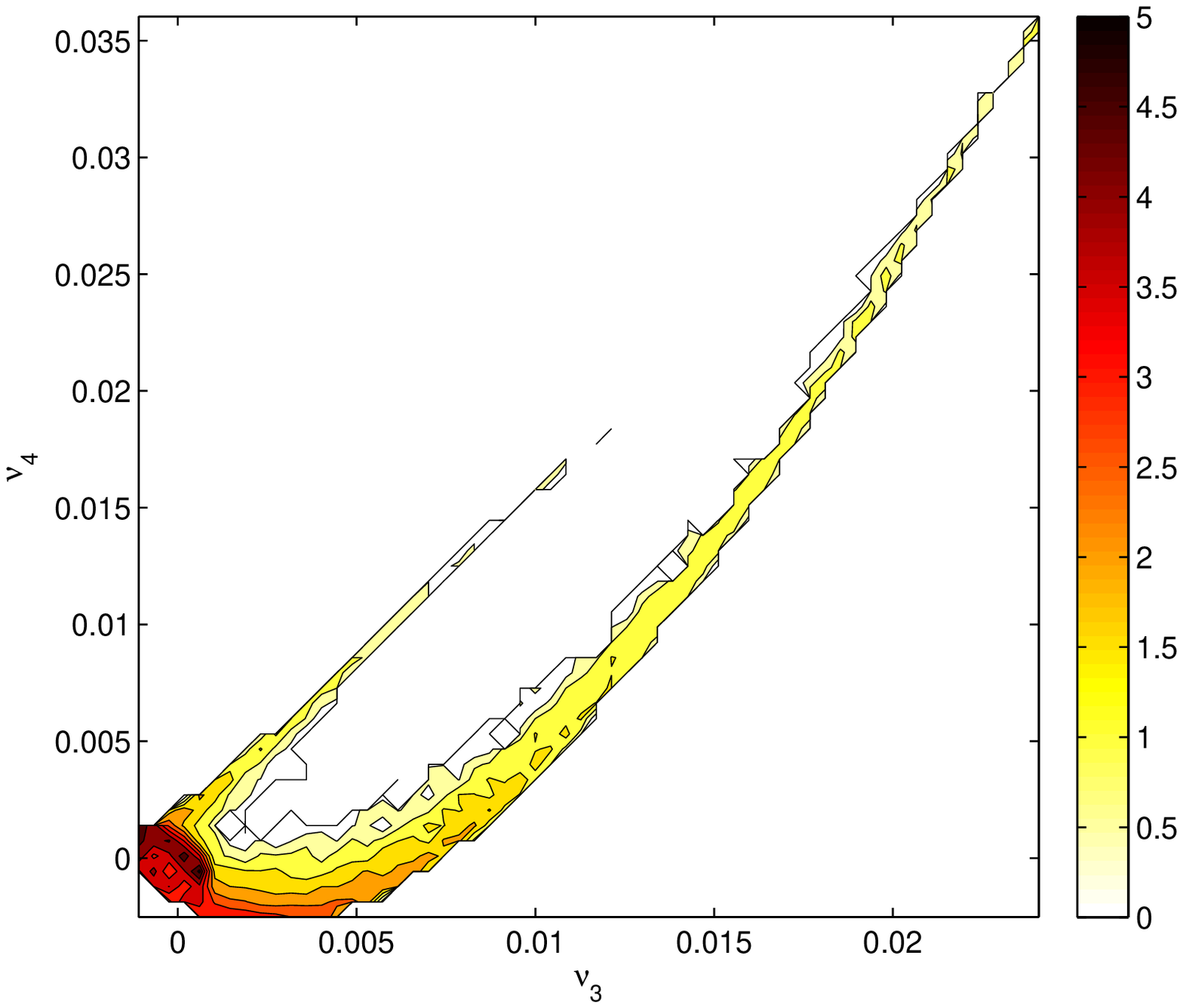}
	 \label{fig:markov-densities-D}
      } 
      \subfigure[]{
         \includegraphics[width=0.30\textwidth]{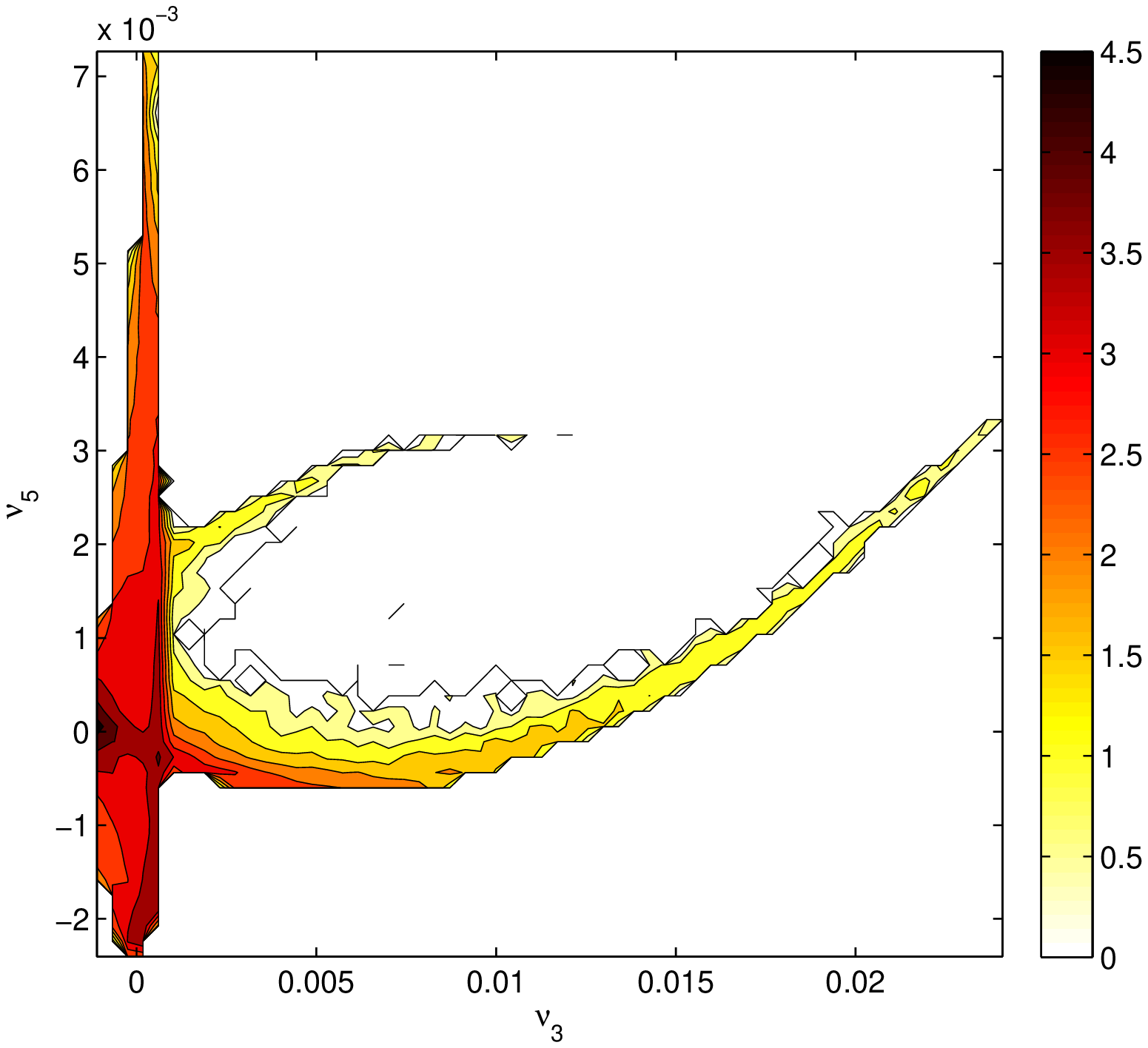}
	 \label{fig:markov-densities-E}
      } 
      \subfigure[]{
         \includegraphics[width=0.30\textwidth]{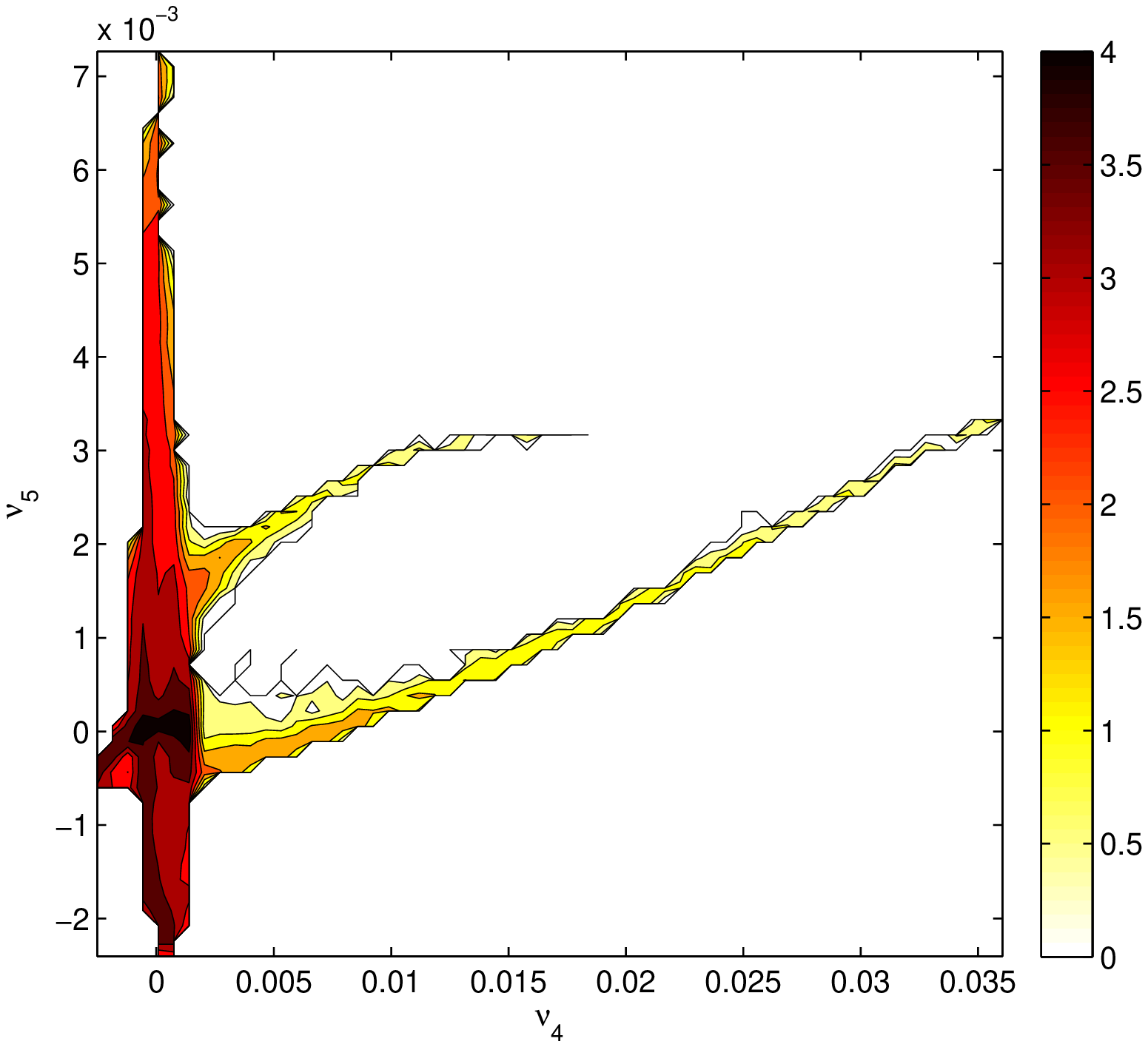}
	 \label{fig:markov-densities-F}
      } 
      \end{center}
\caption{Density (in $\log_{10}$ scale) of galaxies on eigenvectors 2 through 5 of the lazy, autotuned Markov operator: (a) eigenvectors (2,3); (b) eigenvectors (2,4); (c) eigenvectors (2,5); (d) eigenvectors (3,4); (e) eigenvectors (3,5); (f) eigenvectors (4,5).  Plots on all pairs of eigenvectors 2 through 5 are presented as a point of reference for comparison with other figures.}
\label{fig:markov-densities}
\end{figure}

The number density of galaxies on eigenvectors 2 through 5 of the lazy Markov embedding, with $k$=32 and autotuned bandwidths is presented in Figure~\ref{fig:markov-densities}. 
Each of the subfigures presents the same data presented from a different angle---literally, since the data are being projected onto two dimensions of the embedding space.
(We included all of these for comparison with previous and subsequent figures.  For example, 
Figure~\ref{fig:effect-of-k-markov} is taken from the same angle as Figure~\ref{fig:markov-densities-D}, and thus the value of $k$ means that Figure~\ref{fig:effect-of-k-markov-C} corresponds to Figure~\ref{fig:markov-densities-D}.)
Due to the widely varying nature of the density, the color is plotted in logarithmic scale. 
It is clear that the vast majority of galaxies lie on the orange to dark red ``spine" in the lower portion of Figure~\ref{fig:markov-densities-A}, but this appears very different on different pairs of eigenvectors.
For example, the high-density regions are somewhat spread out in Figure~\ref{fig:markov-densities-C}, but they are projected to almost the same point and thus not easily-discriminatable in Figures~\ref{fig:markov-densities-D} and~\ref{fig:markov-densities-F}. 
Figure~\ref{fig:markov-densities-C} also reveals two disjoint regions whose density are an order of magnitude greater than surrounding regions. 
%
As young galaxies evolve and burn up their fuel, they move from the left-most tip of the figure toward the high-density regions as they get redder. 
Yet this subfigure shows that the red galaxies on the right-most ridge are not a ``well-connected'' cluster, in the sense that there is a density minimum between the two peaks. 
This suggests that one can obtain improved quantitative insight about this class of galaxies by studying these embeddings using stellar population synthesis models, a topic which we leave to future work.

\subsection{Continuum Shape and Line Strength}
\label{subsec:embedding-2vs4}

\begin{figure}
      \begin{center}
      \subfigure[]{
         \includegraphics[width=.44\textwidth]{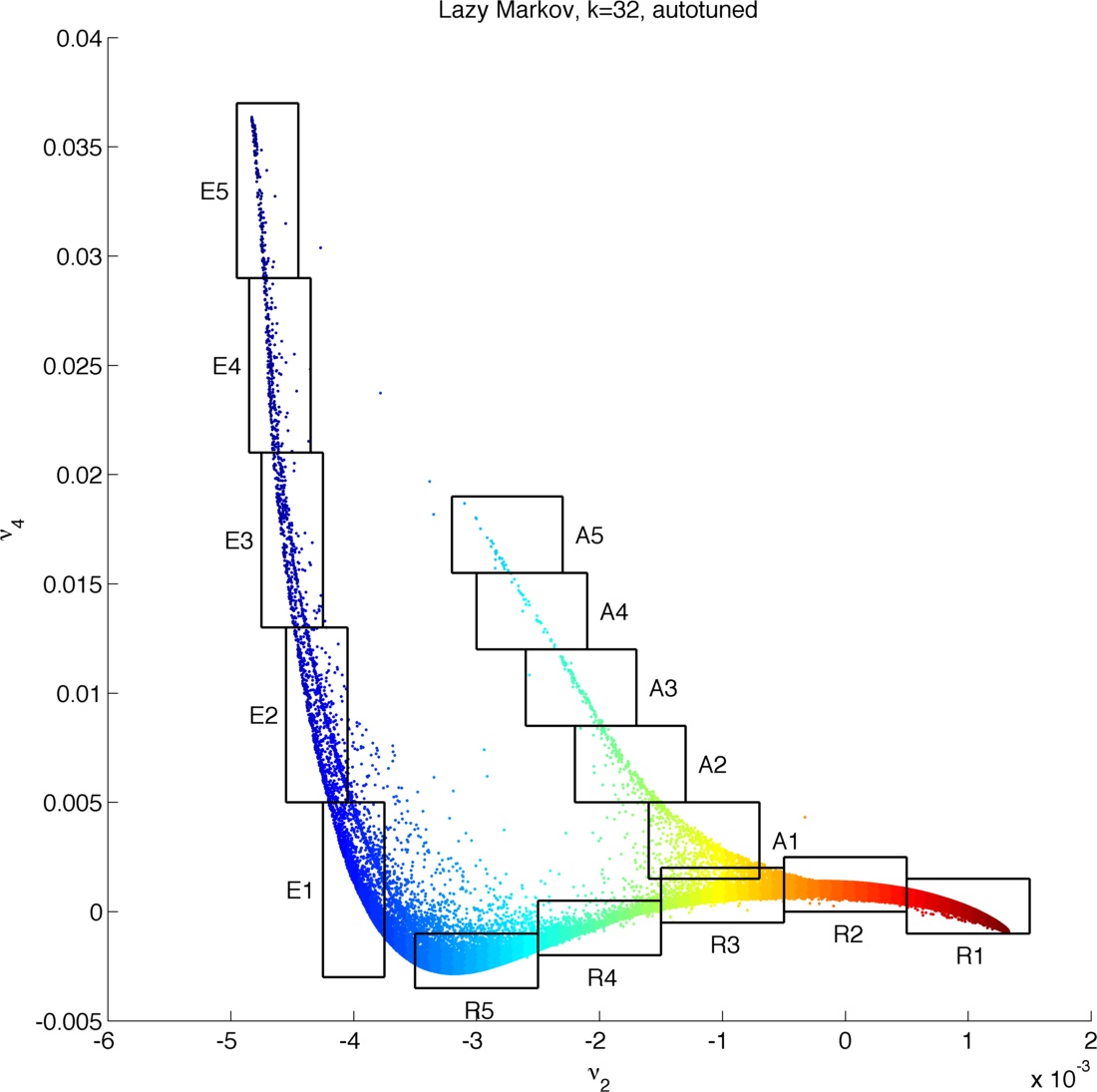}
	 \label{fig:markov-2vs4-reaboxes-A}
      } 
      \subfigure[]{
         \includegraphics[width=.44\textwidth]{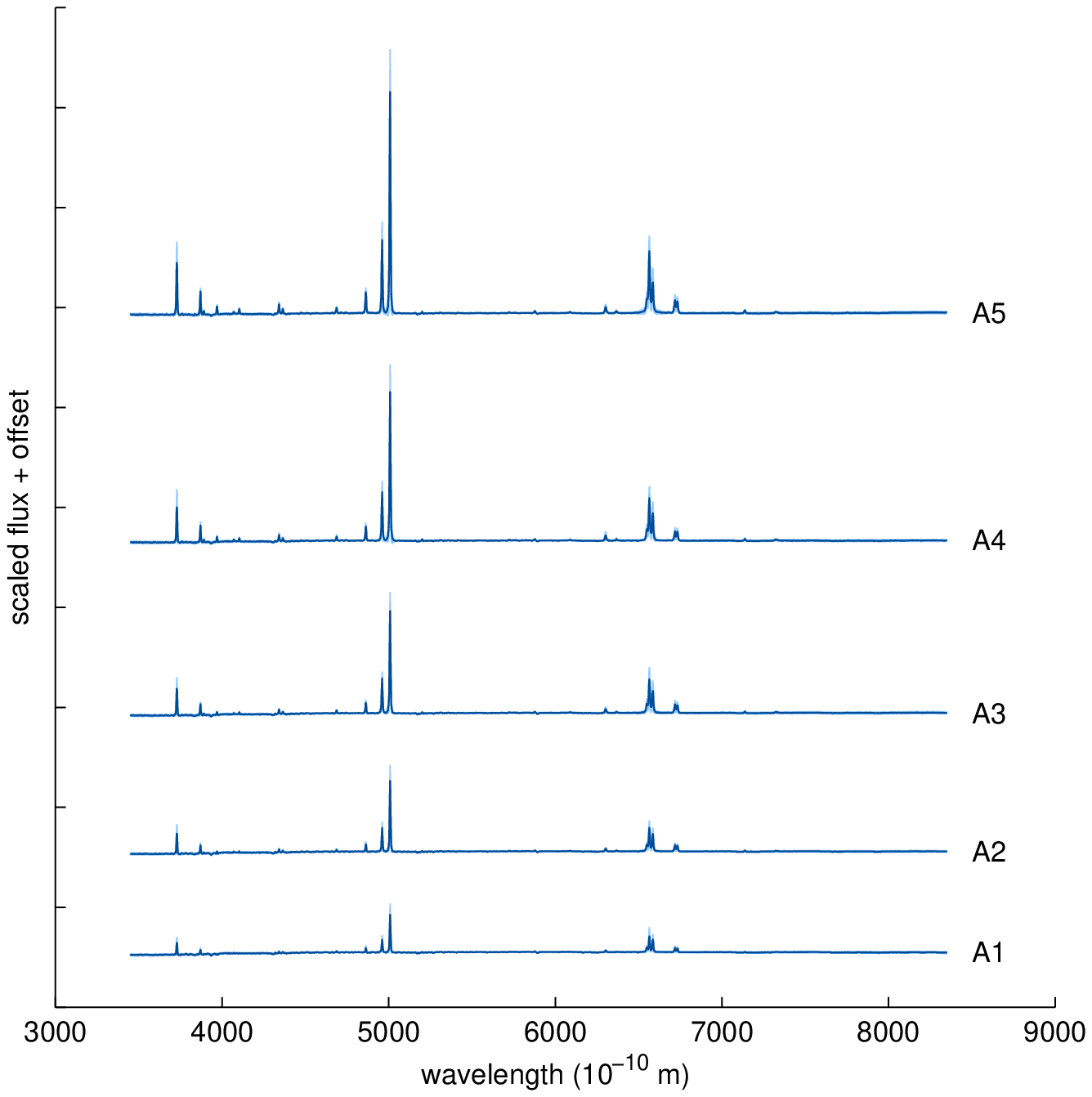}
	 \label{fig:markov-2vs4-reaboxes-B}
      } \\
      \subfigure[]{
         \includegraphics[width=.44\textwidth]{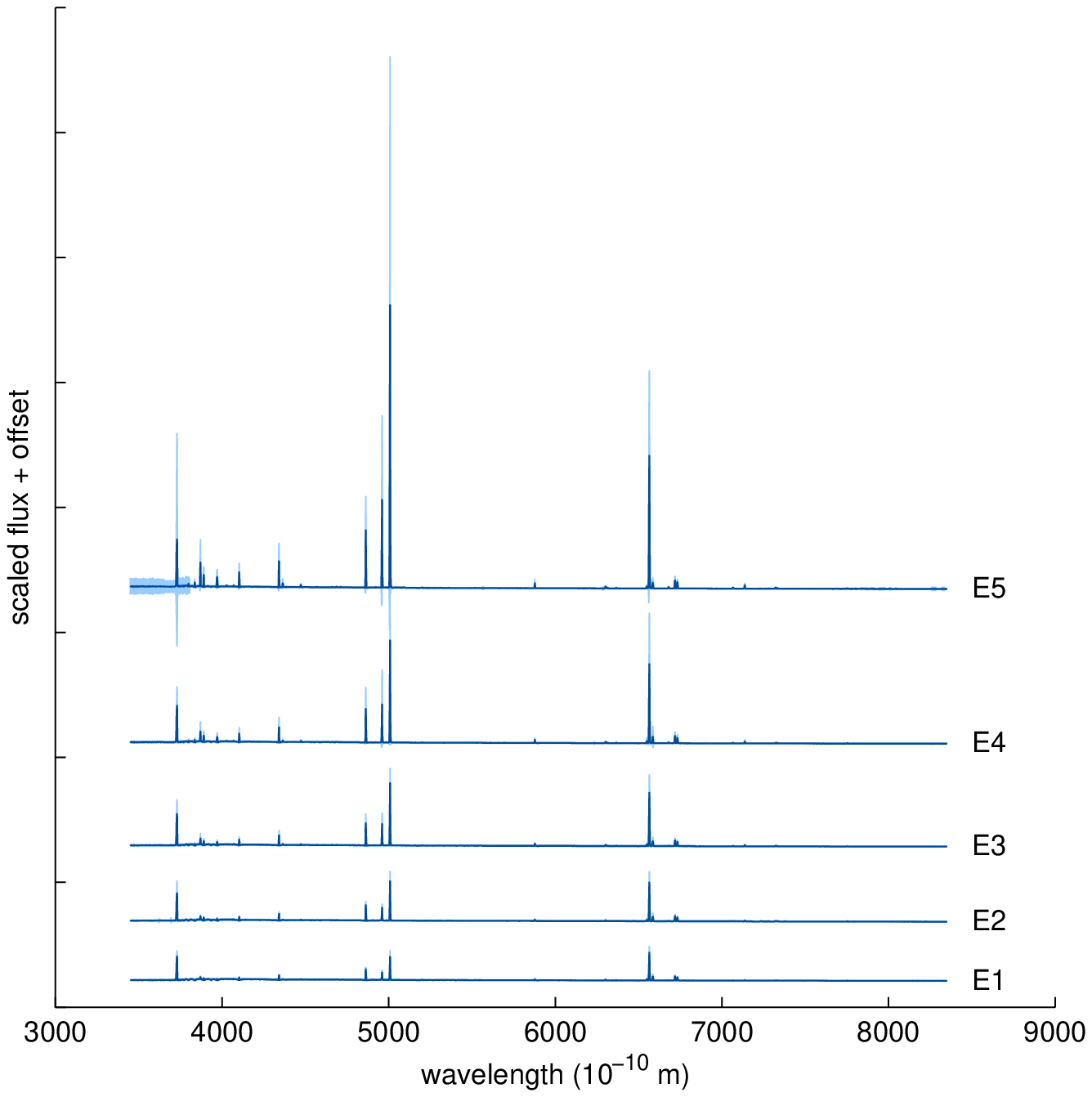}
	 \label{fig:markov-2vs4-reaboxes-C}
      } 
      \subfigure[]{
         \includegraphics[width=.44\textwidth]{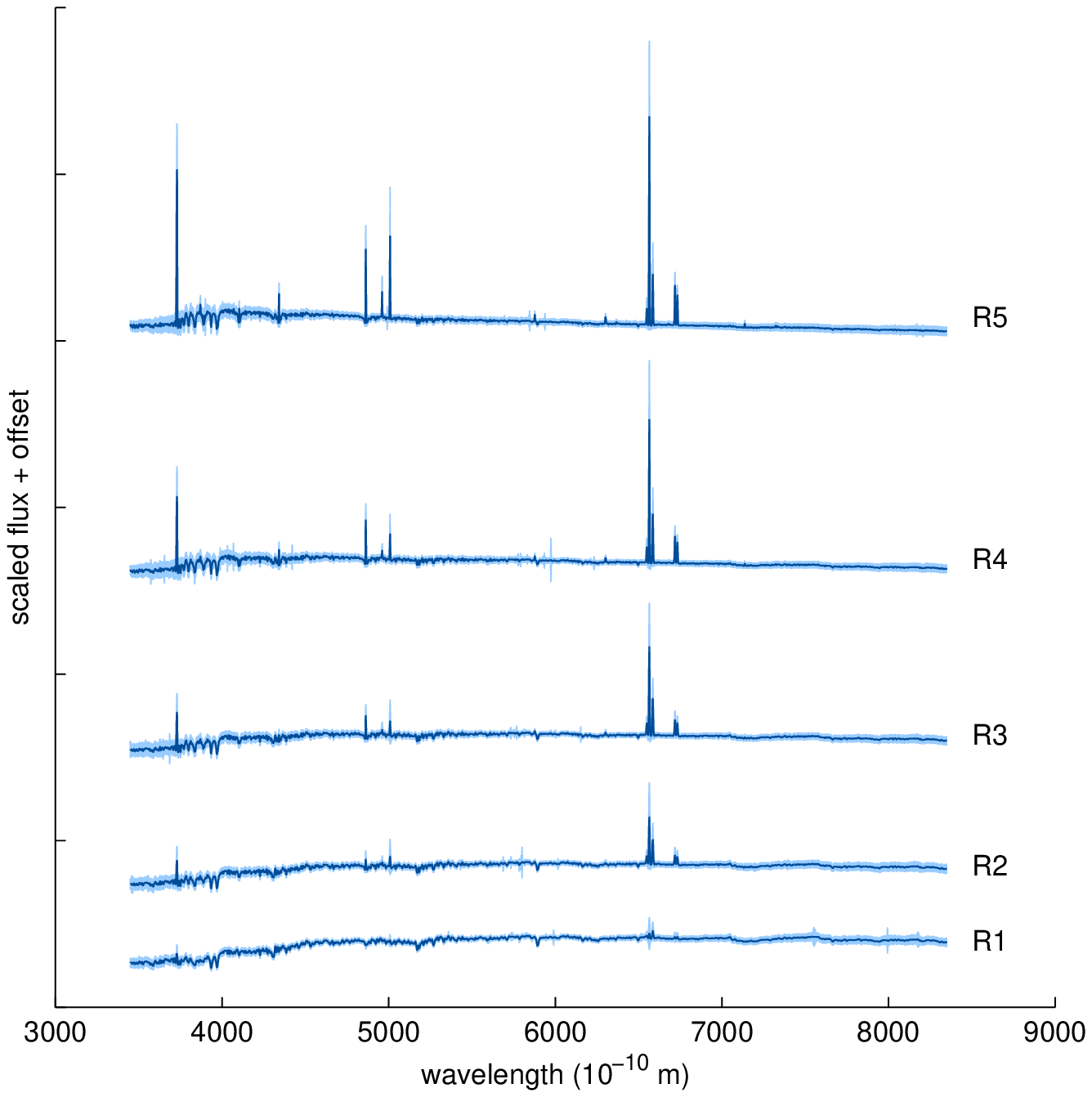}
	 \label{fig:markov-2vs4-reaboxes-D}
      } 
      \end{center}
\caption{(a) Subpopulations of galaxies sampled according to the embedding on eigenvectors 2 and 4 of the Markov operator. (b)-(d) Mean spectrum (in dark blue) and standard deviation (in light blue) for each of the subpopulations in (a).}
\label{fig:markov-2vs4-reaboxes}
\end{figure}

Figure~\ref{fig:markov-2vs4-reaboxes} illustrates how our method can be used to gain insight into the continuum shape and line strength.
To do so, we present a view of the data set as embedded on the second and fourth eigenvectors of $M$ in Figure~\ref{fig:markov-2vs4-reaboxes-A}. 
(Thus, this perspective corresponds to Figure~\ref{fig:markov-densities-B}.)
In this plot, color corresponds to the value of the second eigenvector. 
The boxes labeled A1--A5, R1--R5, and E1--E5 denote regions of embedding coordinates over which we calculated the mean spectra to increase the signal-to-noise ratio.
The corresponding spectra are shown in Figure~\ref{fig:markov-2vs4-reaboxes-B} through~\ref{fig:markov-2vs4-reaboxes-D}.
These delineations were drawn in an ad-hoc manner, and are meant only to guide our interpretation of the embedding dimensions.

In Figure~\ref{fig:markov-2vs4-reaboxes-D}, in particular, we present the mean spectrum (in dark blue) and the standard deviation of the mean (in light blue) of the spectra in the boxes labeled R1--R5 in Figure~\ref{fig:markov-2vs4-reaboxes-A}. 
From these average spectra, we can clearly see the correlation of the second eigenvector with the shape of the continuum: positive values of $\nu_2$ correspond to red continuum shapes, while negative values correspond to blue continuum shapes. 
There is also a clear correlation with the strengths of emissions lines, with negative $\nu_2$ values corresponding to larger fluxes. 
As the continuum spectrum gets bluer, the spectral lines appear in the early-type galaxy indicating star formation. 
A natural continuation of this trend is in the E1--E5 boxes shown in the Figure~\ref{fig:markov-2vs4-reaboxes-C}, where the lines become overwhelmingly strong in the young galaxies.
At first glance, we see a dramatic change in not only the lines but also their ratios. For example, the [O\,{\sc iii}] (4959\AA, 5007\AA) lines grow significantly in comparison to the H\,$\alpha$ (6563\AA). 
In Figure~\ref{fig:markov-2vs4-reaboxes-B}, we present the mean and standard deviation of spectra in boxes A1--A5 of Figure~\ref{fig:markov-2vs4-reaboxes-A}. 
These spectra trace a trend that is different from the main direction, but we see increasing line strengths and bluer spectra. 
%
%
This population of galaxies will be immediately obvious in Section~\ref{subsec:embedding-class}, where we show the results of the SDSS classification \cite{brinchmann04}: these are Active Galactic Nuclei (AGN).

\subsection{A View into Red Ellipticals}
\label{subsec:embedding-2vs5}

\begin{figure}
      \begin{center}
      \subfigure[]{
         \includegraphics[width=.44\textwidth]{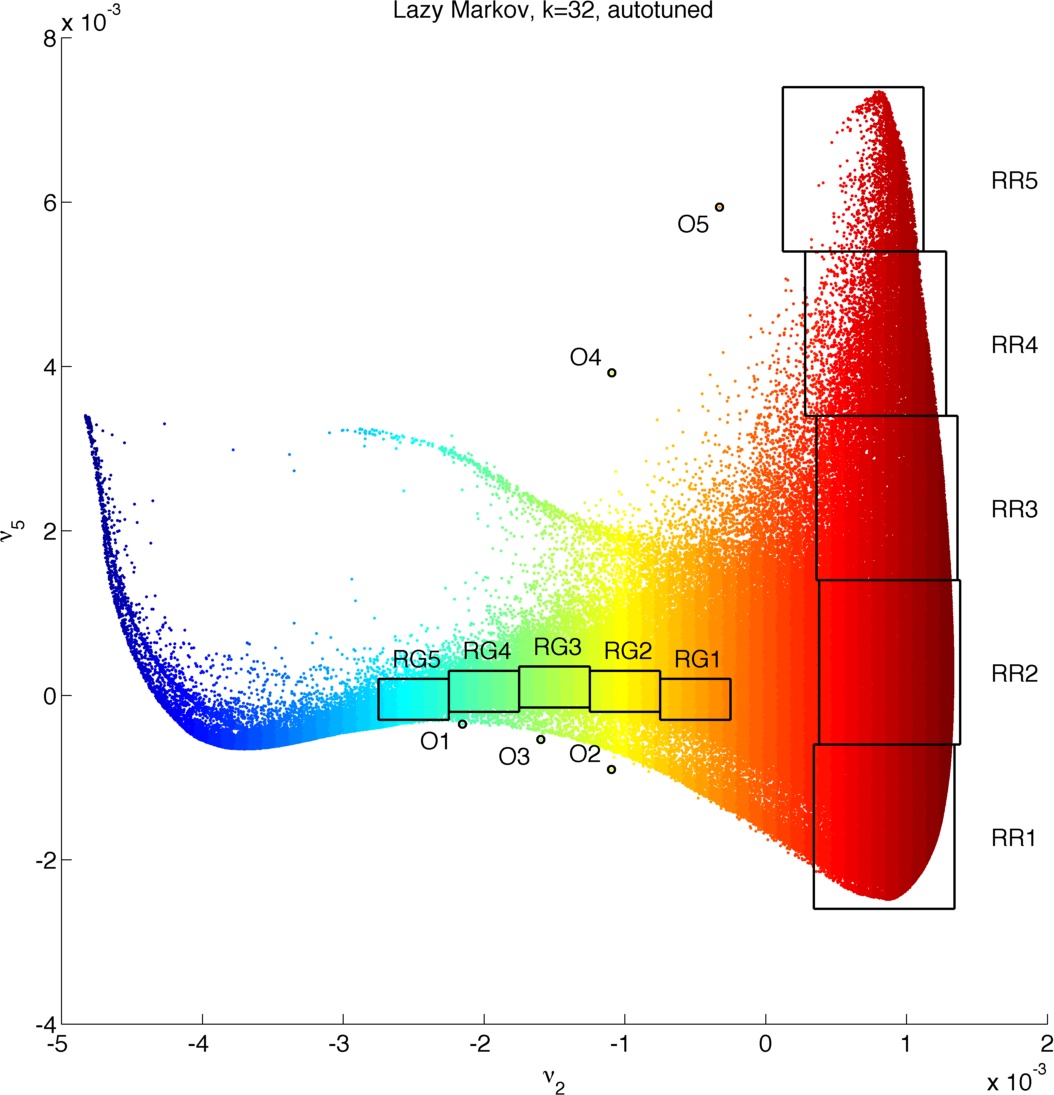}
	 \label{fig:markov-2vs5-rrboxes-A}
      } 
      \subfigure[]{
         \includegraphics[width=.44\textwidth]{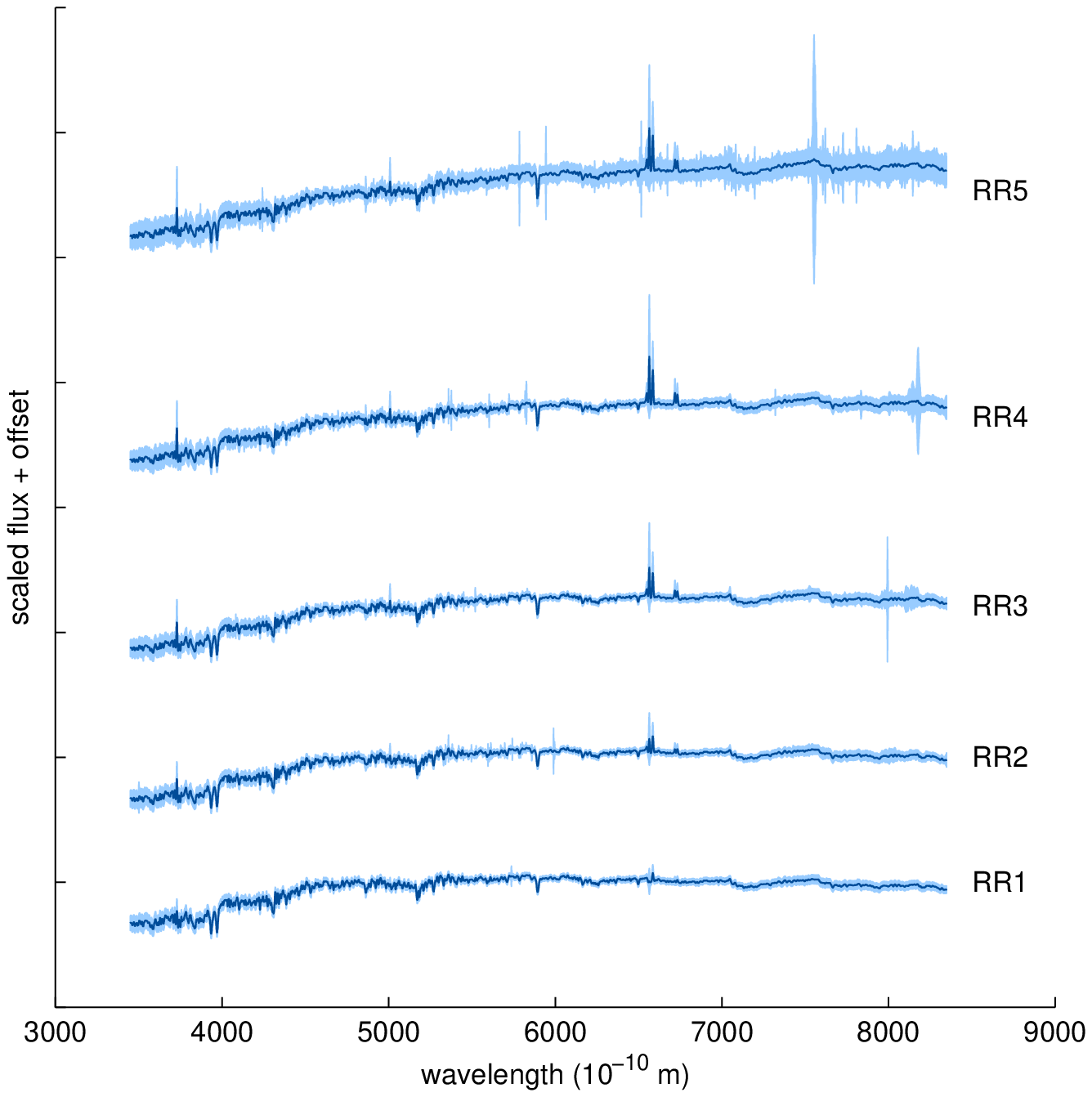}
	 \label{fig:markov-2vs5-rrboxes-B}
      } \\ 
      \subfigure[]{
         \includegraphics[width=.44\textwidth]{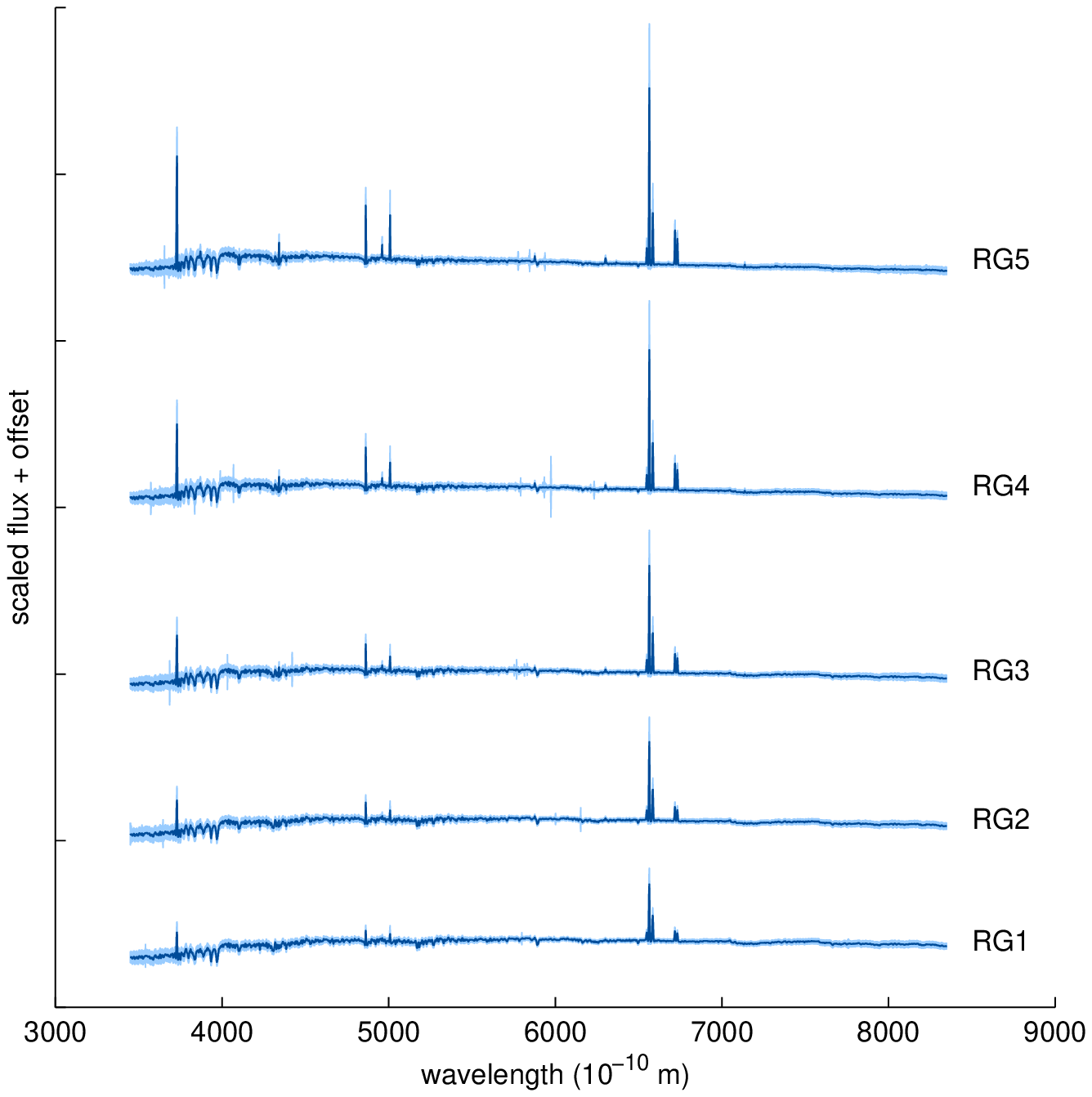}
	 \label{fig:markov-2vs5-rrboxes-C}
      } 
      \subfigure[]{
         \includegraphics[width=.44\textwidth]{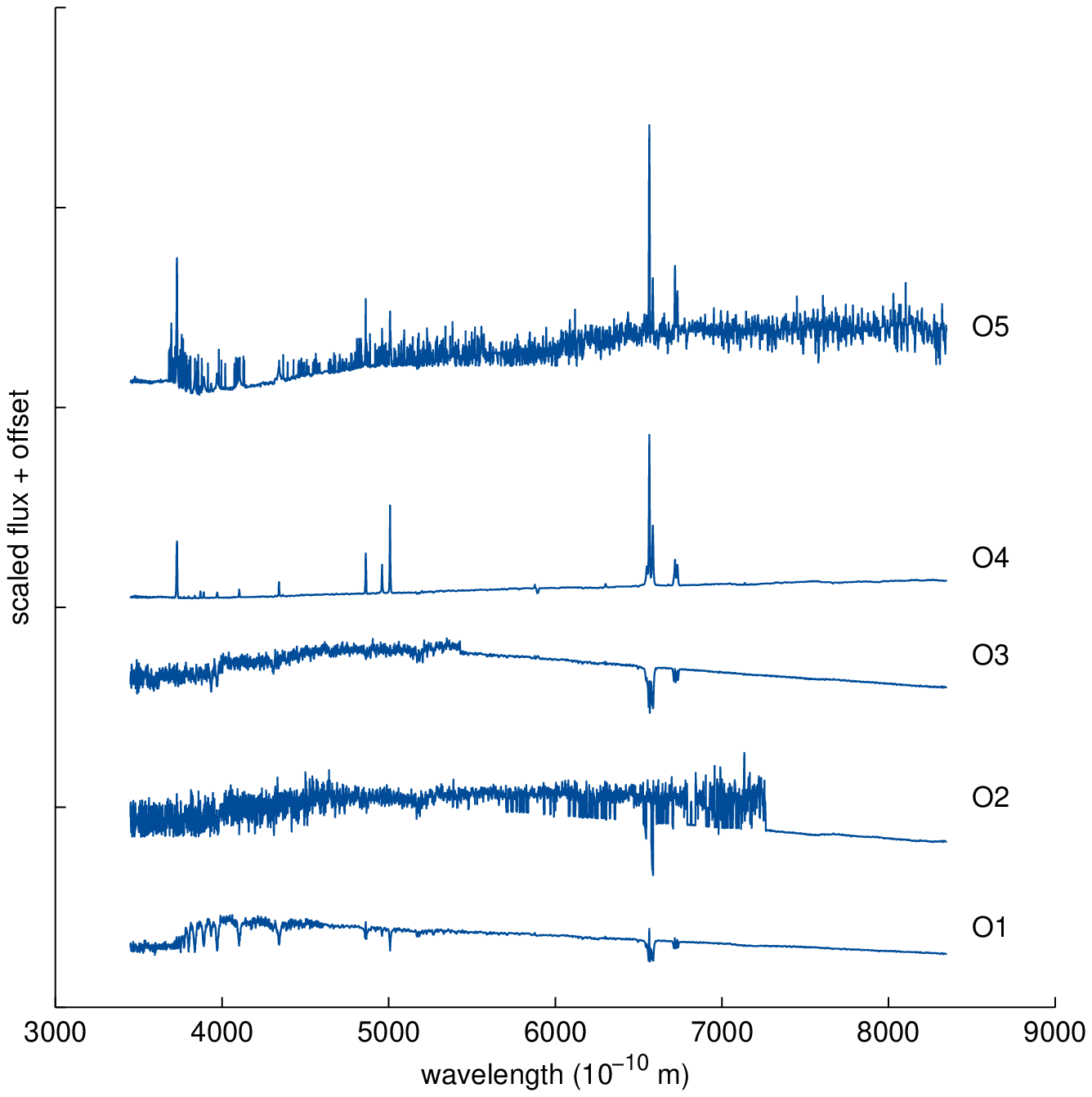}
	 \label{fig:markov-2vs5-rrboxes-D}
      } 
      \end{center}
\caption{(a) Subpopulations of red galaxies sampled according to the embedding on eigenvectors 2 and 5 of the Markov operator. (b)-(c) Mean spectrum (in dark blue) and standard deviation (in light blue) for each of the subpopulations in (a). (d) Selected outlier spectra; O4 has been scaled by a factor of 1/10 and O5 by a factor of 1/100 for legibility.}
\label{fig:markov-2vs5-rrboxes}
\end{figure}

Figure~\ref{fig:markov-2vs5-rrboxes} illustrates how our method can be used to gain insight into the properties of red elliptical galaxies.
In Figure~\ref{fig:markov-2vs5-rrboxes-A} we present a view of the data set as embedded on the second and fifth eigenvectors of $M$. 
(Thus, this perspective corresponds to Figure~\ref{fig:markov-densities-C}.)
Here, we have chosen to explore this projection since $\nu_5$ discriminates red galaxies in the very dense region of spectrum space better than other low-order eigenvectors. 
As in Figure~\ref{fig:markov-2vs4-reaboxes}, color corresponds to the value of $\nu_2$, and we have drawn bounding boxes in an ad-hoc manner.

In Figure~\ref{fig:markov-2vs5-rrboxes-B}, we present the mean and standard deviation of spectra in boxes RR1--RR5 from Figure~\ref{fig:markov-2vs5-rrboxes-A}. 
It is clear that all spectra in this area of the embedding share a red continuum shape, with small or absent emissions lines. 
We also note the increasing strength of H\,$\alpha$ with increasing values of $\nu_5$. 
Comparing with Figure~\ref{fig:markov-densities-C}, it is clear that the density of galaxies also decreases with increasing $\nu_5$, and in particular that box RR1 contains tens of thousands of galaxies.
In Figure~\ref{fig:markov-2vs5-rrboxes-C} we present the mean and standard deviation of spectra in boxes RG1--RG5. 
These correspond to a region of high density, as is evident in Figure~\ref{fig:markov-densities-C}. 
The transition from red to blue continuum shape is perhaps the most evident in this small region of the embedding space.

Finally, for completeness, in Figure~\ref{fig:markov-2vs5-rrboxes-D} we present selected outlier spectra, labeled O1--O5 in Figure~\ref{fig:markov-2vs5-rrboxes-A}. 
Spectra O4 and O5 have been scaled by factors of 1/10 and 1/100, respectively, for legibility. 
It is common to want to identify outliers, either to clean up the data or in the hopes of identifying new phenomena.
In this case, all of these spectra that were identified as outlying by our method appear to be artifacts or errors in the pre-processing, e.g., the gap correction. 
We note that each of these erroneous spectra appear separated from the remainder of the data, indicating the robustness and usefulness of the method for identifying outliers. 
These artifacts are truly exceptions and have been included for illustrative purposes.

\subsection{Relationship to \texttt{eCoeff}}
\label{subsec:pca-eigs}

Here, we compare our embeddings with those obtained via PCA, a dimension-reduction method that optimally preserves linear structure in high-dimensional data sets. 
These embeddings are computed in the SDSS pipeline and stored as \texttt{ecoeff\_i} for $\texttt{i}$ ranging from 0 to 4. 
(The data are not mean-centered in this computation, so \texttt{ecoeff\_0} corresponds to the projection onto the average spectrum.) 
Recall that the coefficients of the expansion on the eigenspectra have previously been used to classify galaxy types~\cite{yip2004distributions}, among many other uses.

\begin{figure}
      \begin{center}
      \subfigure[]{
         \includegraphics[width=.44\textwidth]{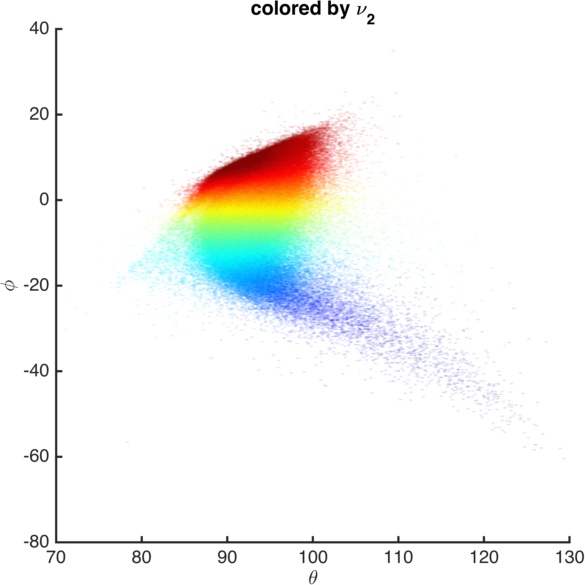}
	 \label{fig:pca-eigs-A}
      } 
      \subfigure[]{
         \includegraphics[width=.44\textwidth]{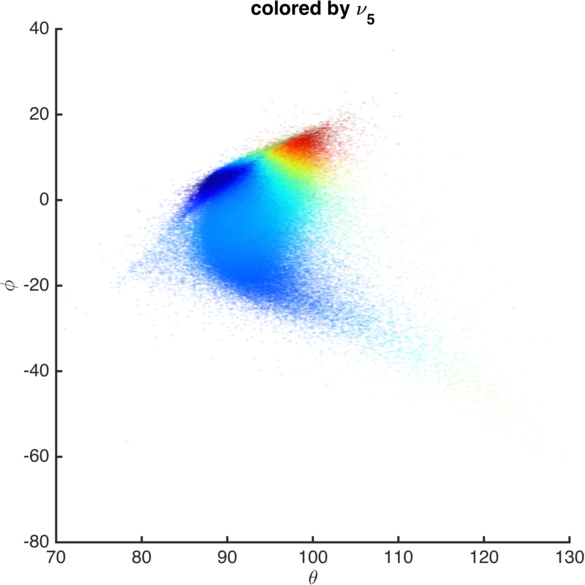}
	 \label{fig:pca-eigs-B}
      } 
      \end{center}
\caption{Embedding of galaxies on the PCA mixing angles $\theta$ and $\phi$, color-coded by eigenvectors 2 and 5 of the lazy, autotuned, $k=32$ Markov operator.} 
\label{fig:pca-eigs}
\end{figure}

In Figure~\ref{fig:pca-eigs}, we plot the galaxies in the embedding on the mixing angles $\theta$ and $\phi$, where in this figure the opacity is proportional to the density and the coloring is determined by eigenvector 2 (in Figure~\ref{fig:pca-eigs-A}) or eigenvector 5 (in Figure~\ref{fig:pca-eigs-B}) of the lazy, autotuned, $k = 32$ Markov operator. 
From Figure~\ref{fig:pca-eigs-A} it is clear that $\nu_2$ and $\phi$ are highly correlated. 
Given our results in the previous subsection, this is not surprising, since both measures mediate between red and blue continuum shapes. 
Not shown are figures showing that $\nu_3$ and $\nu_4$ discriminating among blue spectra, while Figure~\ref{fig:pca-eigs-B} displays how $\nu_5$ picks up in a more complicated way variation among red spectra.

\subsection{Relationship with BPT Diagrams}
\label{subsec:bpt}

\begin{figure}
      \begin{center}
      \subfigure[]{
         \includegraphics[width=.30\textwidth]{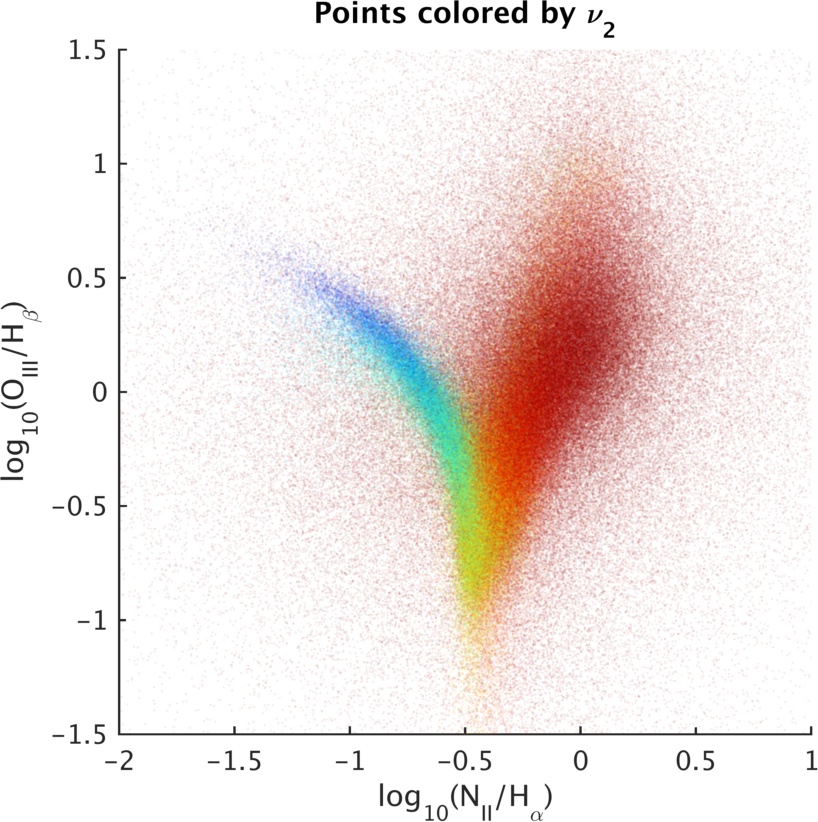}
         \label{fig:bpt-eigs-A}
      } 
      \subfigure[]{
         \includegraphics[width=.30\textwidth]{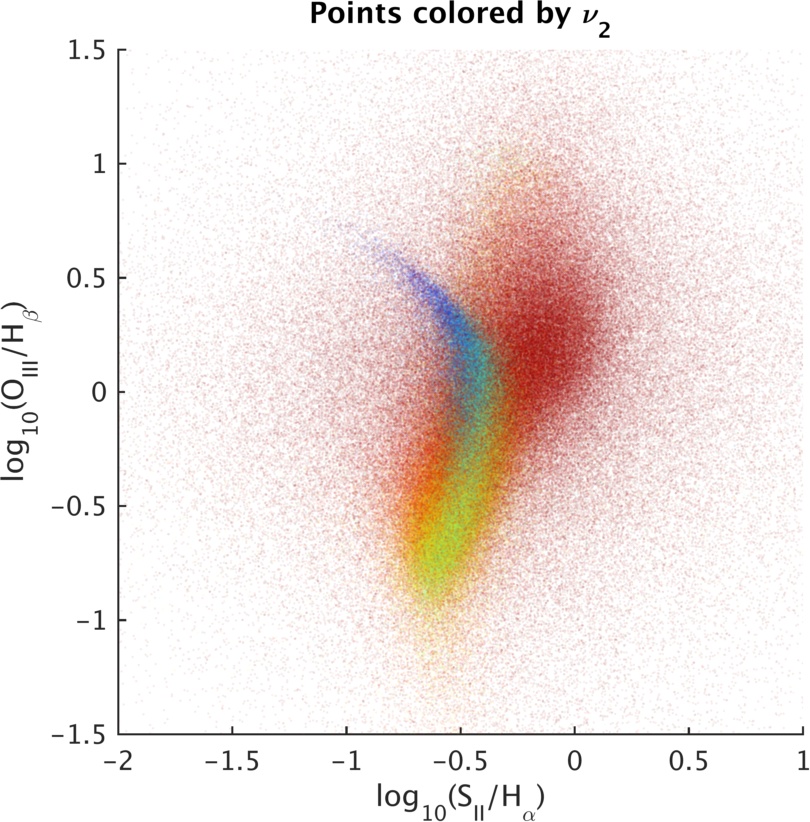}
         \label{fig:bpt-eigs-B}
      } 
      \subfigure[]{
         \includegraphics[width=.30\textwidth]{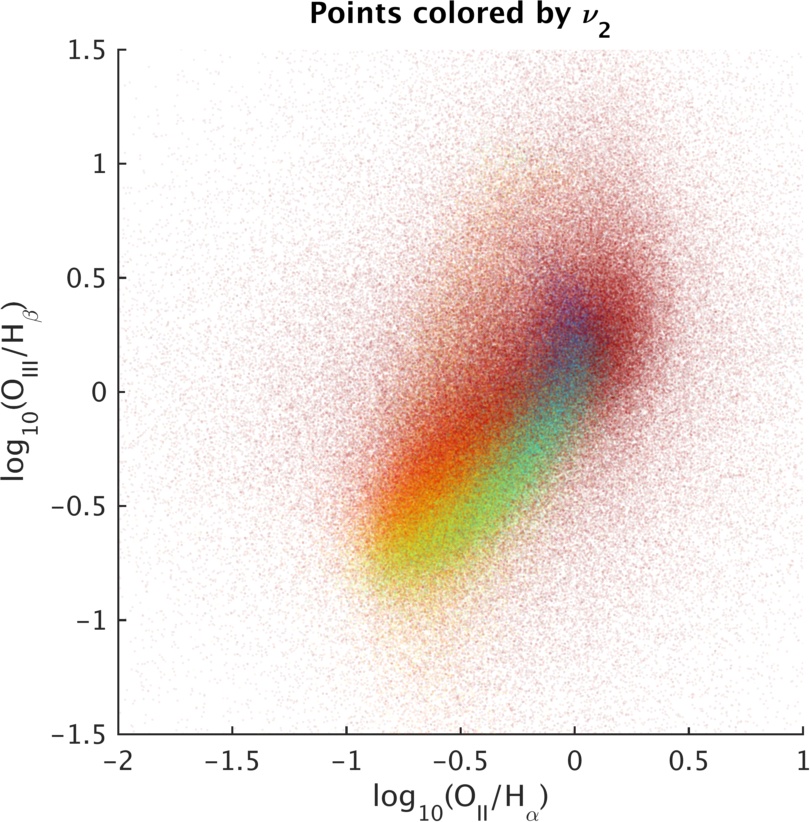}
         \label{fig:bpt-eigs-C}
      } 
      \end{center}
\caption{BPT line-ratio diagrams, color-coded by second eigenvector of the lazy, autotuned, $k=32$ Markov operator.}
\label{fig:bpt-eigs}
\end{figure}

Here, we compare our embeddings with those based on certain emission line strength ratios, known as BPT diagrams. 
These line-ratio embeddings are based on the flux in four wavelength bins, corresponding to the $\mathrm{N_{II},\, H_{\alpha},\, O_{III}}$, and $\mathrm{H_{\beta}}$ emissions lines. 
In Figure~\ref{fig:bpt-eigs}, we show the embedding plot the galaxies in the embedding on the ratios 
\begin{equation}
	\label{eq:bpt-ratios}
	\log_{10}\left( \frac{\mathrm{N_{II}}}{\mathrm{H\,{\alpha}}} \right), \quad
	\log_{10}\left( \frac{\mathrm{O_{III}}}{\mathrm{H\,{\beta}}}\right), \quad \text{and} \quad
	\log_{10}\left( \frac{\mathrm{O_{II}}}{\mathrm{H\,{\beta}}}\right)   ,
\end{equation}
and as in Figure~\ref{fig:pca-eigs} the opacity is proportional to the density and the coloring is determined by the second eigenvector of the lazy, autotuned, $k=32$ Markov operator.
These color versions of Figure~\ref{fig:intro-embeddings-bpt} show that the apparent bifurcation in the BPT plots is really a continuous and gradual change. 
The most dominant component shown in color resolves the degeneracies in these plots and shows that the scatter is primarily due to the red galaxies. 
These plots hint at the possibility of a better classification algorithm that uses the coordinates provided by our method instead of just the BPT line measurements, and this is an obvious direction for future work.

\subsection{Comparison to the SDSS Classification}
\label{subsec:embedding-class}

\begin{figure}
      \begin{center}
         \includegraphics[width=.44\textwidth]{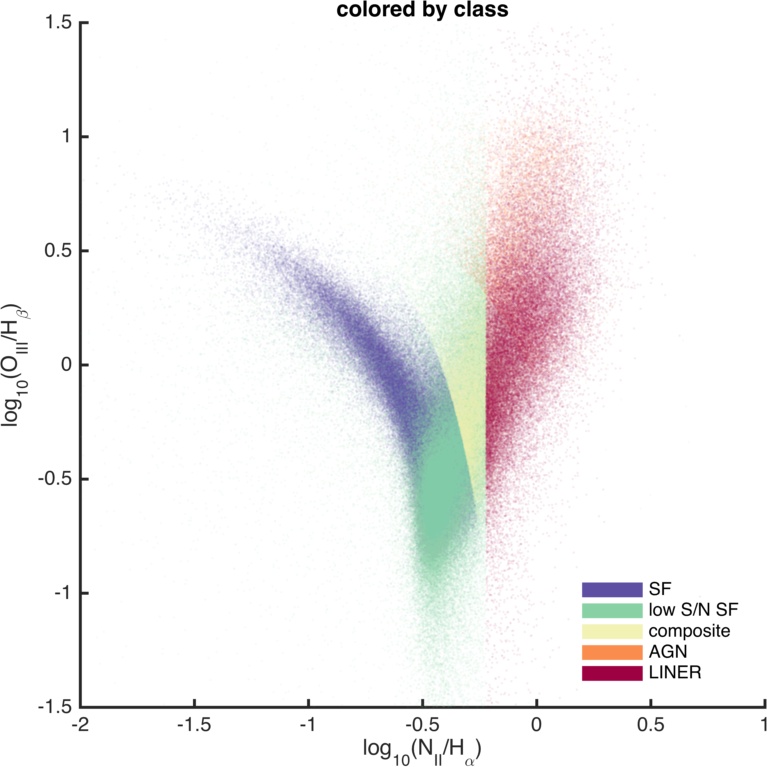}
      \end{center}
\caption{BPT line-ratio diagram of the SDSS DR7 MGS, with galaxies color-coded by class label.}
\label{fig:bpt-eigs-classlabel}
\end{figure}

Here, we augment our embeddings with the addition of the SDSS class labels as derived by \cite{brinchmann04}.  
Our sample of about $517,000$ spectra are split into the following six categories: star-forming; low signal-to-noise star-forming; composite; active galactic nuclei (AGN); low signal-to-noise LINER; and unclassified. 
This state-of-the-art classification scheme goes beyond the BPT diagrams and uses a total of 7 lines to distinguish the separate classes.
See Figure~\ref{fig:bpt-eigs-classlabel} for BPT line-ratio diagrams with galaxies color-coded by class label.
(That is, the color-coding in this figure depends on these galaxy labels and is different than that used in previous~figures.)
The linear class boundaries through high-density regions of the data, as well as a quick comparison with Figure~\ref{fig:bpt-eigs-A}, highlights the arbitrariness of the class boundaries.

\begin{figure}
      \begin{center}
      \subfigure[]{
         \includegraphics[width=0.30\textwidth]{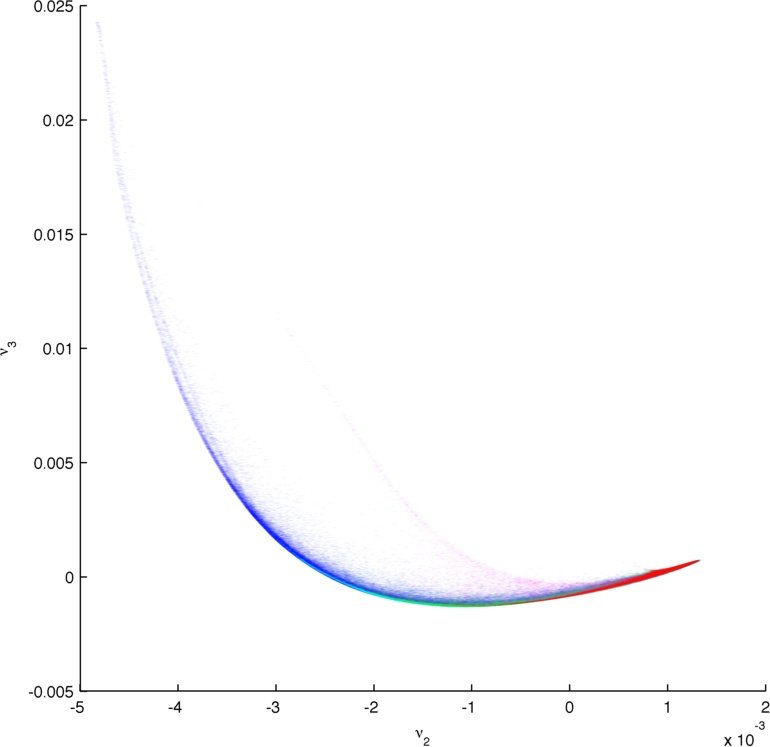}
         \label{fig:global-class-label-A}
      } 
      \subfigure[]{
         \includegraphics[width=0.30\textwidth]{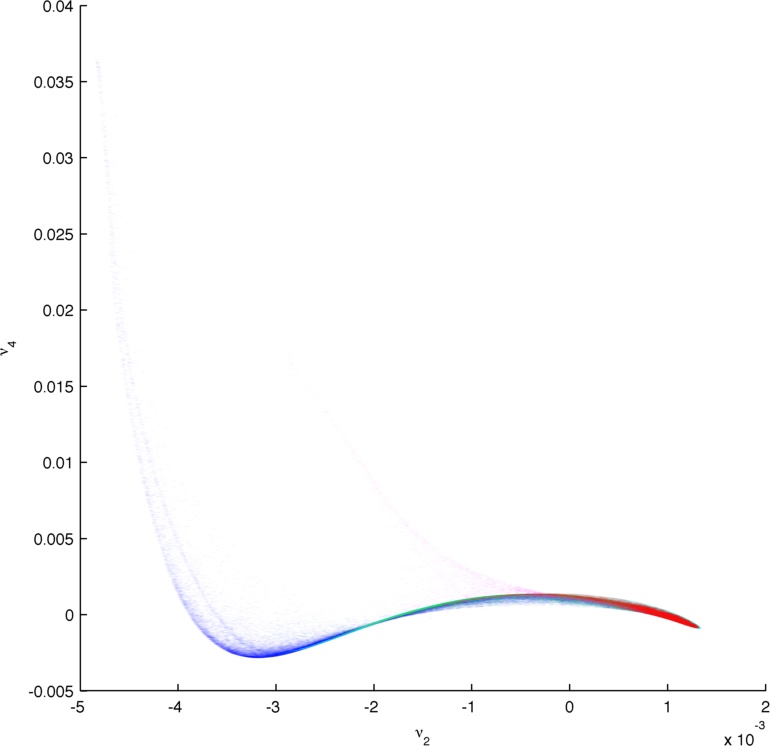}
         \label{fig:global-class-label-B}
      } 
      \subfigure[]{
         \includegraphics[width=0.30\textwidth]{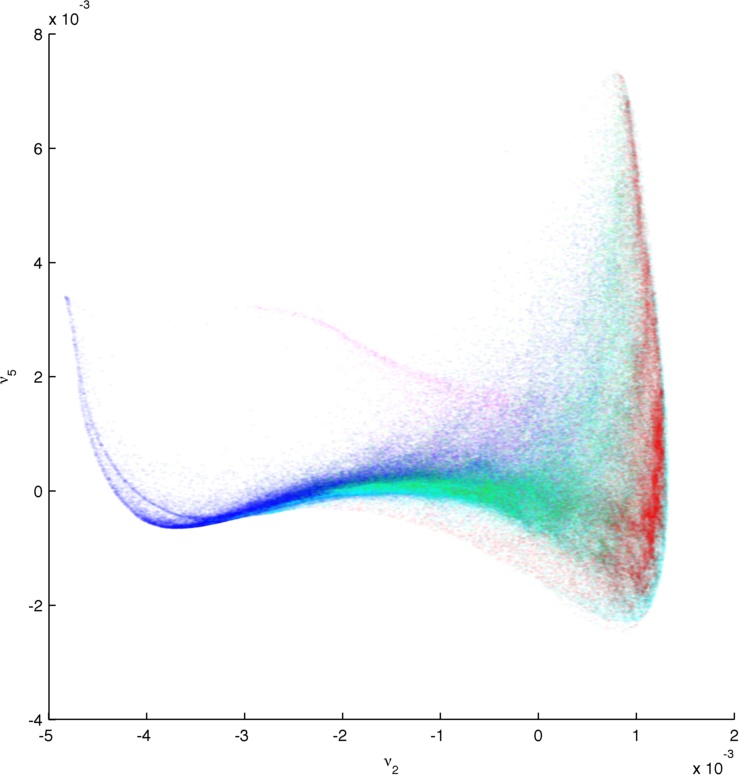}
         \label{fig:global-class-label-C}
      } 
      \subfigure[]{
         \includegraphics[width=0.30\textwidth]{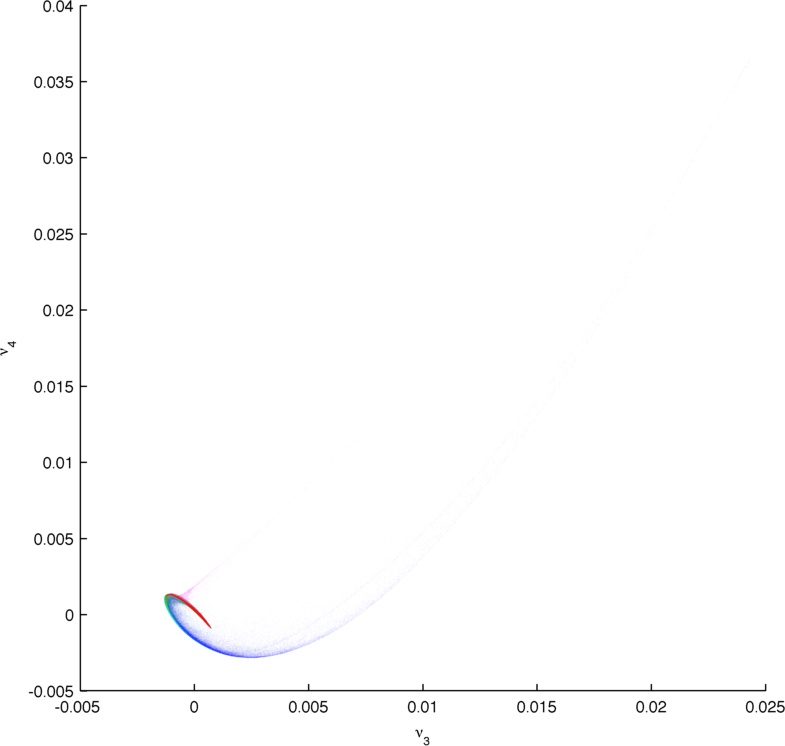}
         \label{fig:global-class-label-D}
      } 
      \subfigure[]{
         \includegraphics[width=0.30\textwidth]{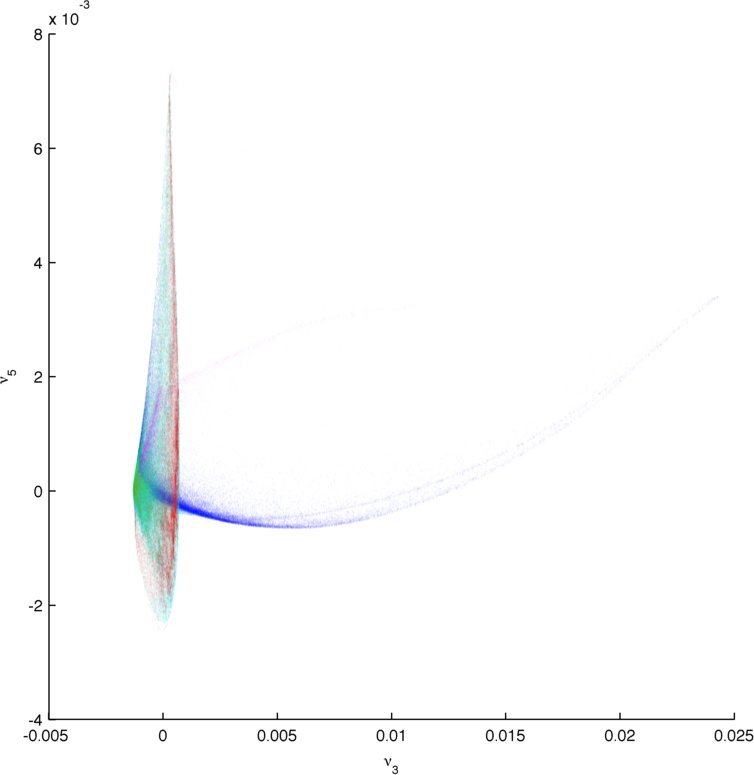}
         \label{fig:global-class-label-E}
      } 
      \subfigure[]{
         \includegraphics[width=0.30\textwidth]{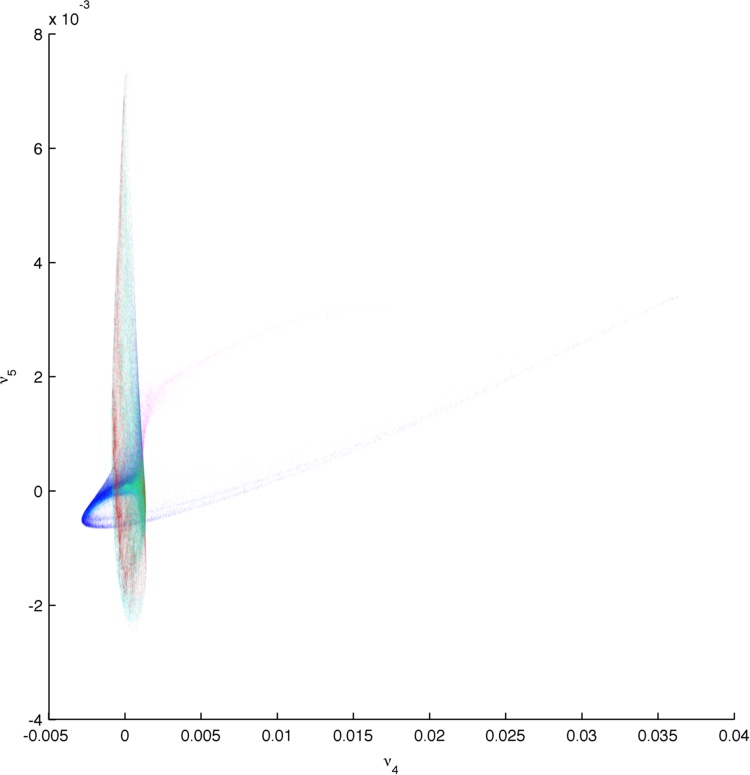}
         \label{fig:global-class-label-F}
      } 
      \end{center}
\caption{Top four nontrivial global eigenvectors of the lazy, autotuned Markov operator, color-coded by type: blue for star-forming; cyan for low signal-to-noise star-forming; green for composite; magenta for active galactic nuclei (AGN); and red for LINER. Unclassified spectra are omitted from this figure. (a) Embedding on eigenvectors (2,3); (b) embedding on eigenvectors (2,4); (c) embedding on eigenvectors (2,5); (d) embedding on eigenvectors (3,4); (e) embedding on eigenvectors (3,5); (f) embedding on eigenvectors (4,5).  For comparison, the ordering of the subfigures is the same as in Figure~\ref{fig:markov-densities}, but the different color-coding and handling of the density makes these subfigures look different.}
\label{fig:global-class-label}
\end{figure}

In Figure~\ref{fig:global-class-label}, we present the embeddings on eigenvectors 2 through 5 of the lazy, autotuned Markov operator with $k=32$. 
We have color-coded the points according to their type: blue for star-forming, cyan for low signal-to-noise star-forming, green for composite, magenta for AGN, and red for LINER. 
We have omitted unclassified spectra from these figures in order to make the embeddings more legible; and we have adjusted the transparency of the plotted points so that the density of spectra is visually evident.
In Figure~\ref{fig:global-class-label-A}, one can discern a clear transition from star-forming to composite to LINER with increasing $\nu_2$. 
This agrees with the earlier remarks regarding the correlation of $\nu_2$ with continuum shape. 
In addition, as noted previously, the AGN form a separate ``spur" between the star-forming galaxies and LINERs. 
In Figure~\ref{fig:global-class-label-C}, we note the concentration of composite galaxies along the high-density spine evident in Figure~\ref{fig:markov-densities-C} that was investigated in Figure~\ref{fig:markov-2vs5-rrboxes-C}. 
These spectra exhibit a clear transition from blue to red continuum shapes, which agrees with their position in our embeddings between star-forming and LINER galaxies. 
We also note the concentration of LINER and low signal-to-noise star-forming galaxies along the lower right rim of the embedding. 
(We will see in Section~\ref{sec:local} that these two types are difficult to distinguish from one another.)
In Figures~\ref{fig:global-class-label-C}--\ref{fig:global-class-label-F}, we note the separation between the bulk of the spectra and those labeled as AGN. 
In particular, in Figure~\ref{fig:global-class-label-D} one can discern a decision boundary for AGN in the purple axis behind the ``fan" of composite and LINER spectra. 
Again, this is striking visual evidence of the arbitrariness of this class boundary. 

\subsection{Bimodality of the Blue Ridge in Figure~\ref{fig:markov-2vs4-reaboxes-A}}
\label{subsec:redshift}

\begin{figure}
      \begin{center}
      \subfigure[]{
         \includegraphics[width=.44\textwidth]{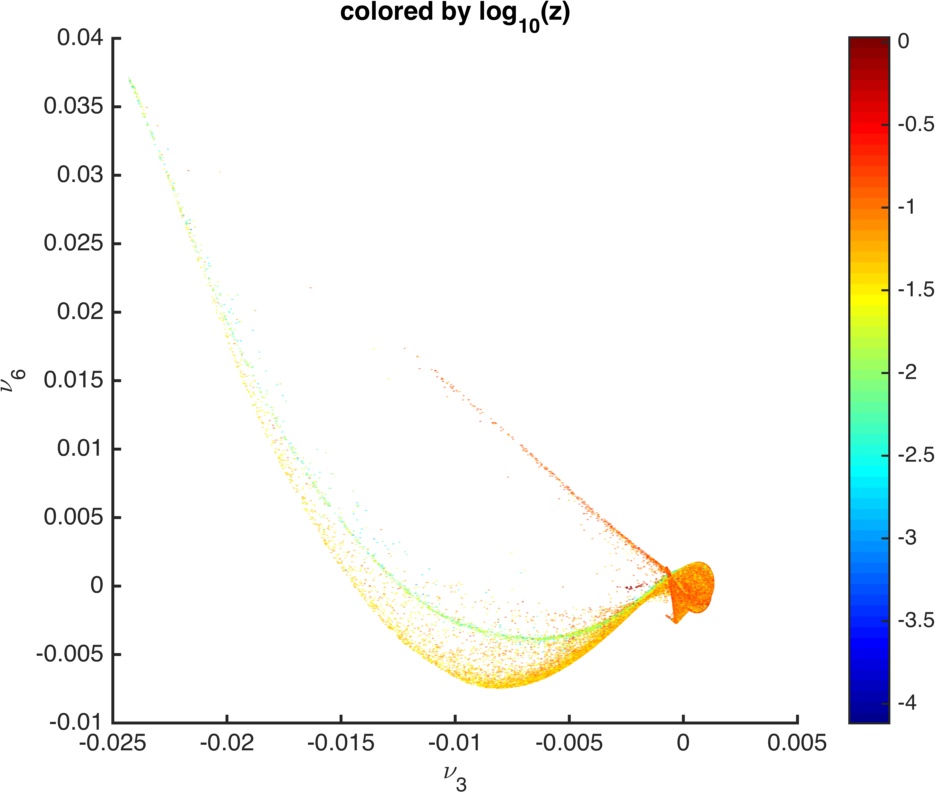}
         \label{fig:redshift-A}
      } 
      \subfigure[]{
         \includegraphics[width=.44\textwidth]{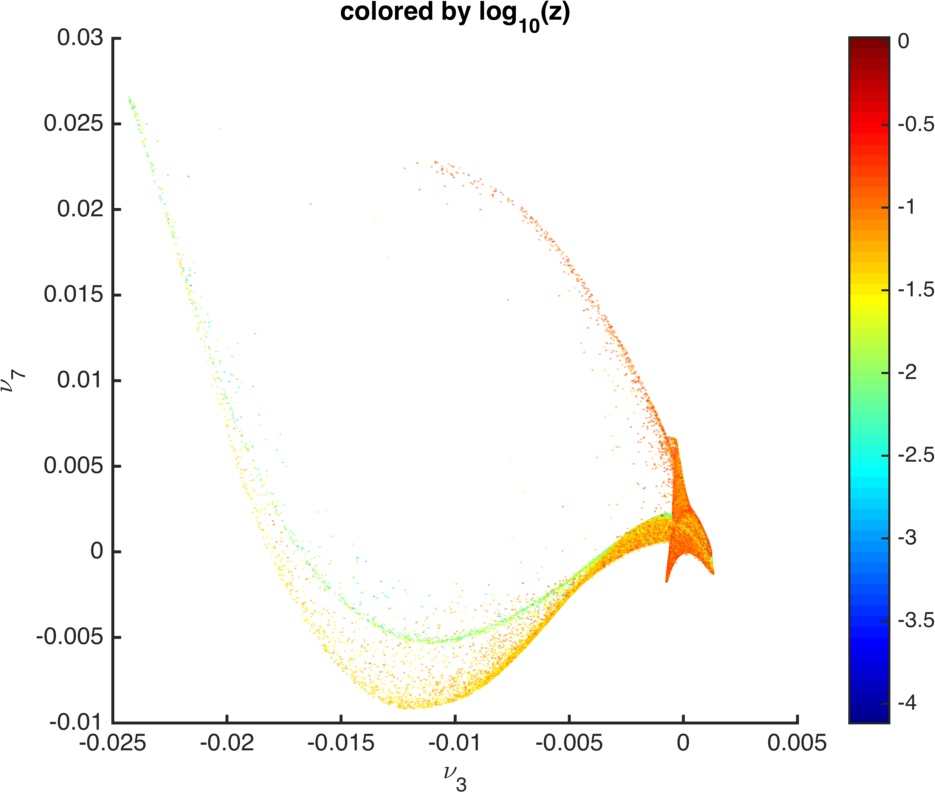}
         \label{fig:redshift-B}
      } 
      \end{center}
\caption{Embeddings of SDSS DR7 MGS on higher-order eigenvectors, 3rd versus 6th and 3rd versus 7th in \ref{fig:redshift-A} and \ref{fig:redshift-B}, respectively, color-coded by redshift in logarithmic scale.}
\label{fig:redshift}
\end{figure}

We conclude this section by noting that, upon closer inspection of Figure~\ref{fig:markov-2vs4-reaboxes-A}, the blue sequence of galaxies actually appear to have a bifurcation and there are two parallel ridges going through the the boxes E1--E3. 
To highlight and understand better this apparent bimodality, consider Figure~\ref{fig:redshift}, where we present multiple embeddings on two pairs of higher-order eigenvectors. 
In these panels, the separate trendline is clearly visible. 
The key insight, however, comes from the color scheme we used: it simply represents the redshift of each galaxy. 
That is, again, the color-coding in Figure~\ref{fig:redshift} is incomparable with that of Figure~\ref{fig:markov-2vs4-reaboxes}, as here it is determined by the redshift of the galaxy.
We see that the one of the strands contains the lowest redshift spectra with $z\!<\!0.02$ that belong to the largest galaxies on the sky, and one might wonder whether this is of astronomical significance. 
Upon examination, we have determined that this is an artifact: the SDSS photo pipeline is known to break apart large galaxies, and here we witness its power to pick out individual star-forming H\,{\sc ii} regions, which the target selection identified as galaxies. 
Using the SkyServer Image Cutout services, we checked the thumbnail of these sources, and their morphology is in complete agreement. 
In some of the cases the enhanced star-formation appears to be induced by merging galaxies.
 
\section{Local Structure via Local Embeddings}
\label{sec:local}

In this section, we provide examples of how our method can be used as an exploratory tool to identify small-scale local structure in the SDSS data.
Previous work has shown performance gains using appropriately-seeded locally-biased semi-supervised eigenvector embeddings for tasks such as classifying handwritten digits and brain activity in fMRI images~\cite{HM14_JRNL}.
Thus, we expect that by seeding our local embeddings in an appropriate manner, we should be able to identify small-scale local properties of the data as well as to extract more information from a smaller number of features, especially as pertains to the classification of smaller classes of~galaxies.

\subsection{Zooming in Locally around Given Seeds}
\label{sec:local-zoom}

\begin{figure}
      \begin{center}
      \subfigure[]{
         \includegraphics[width=.30\textwidth]{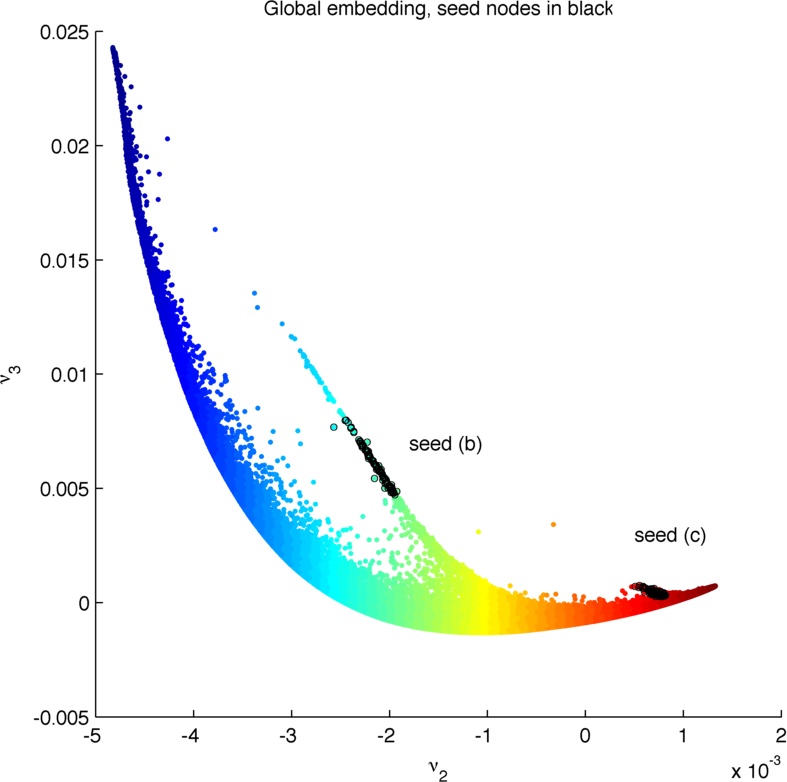}
	 \label{fig:effect-of-seed-A}
      } 
      \subfigure[]{
         \includegraphics[width=.30\textwidth]{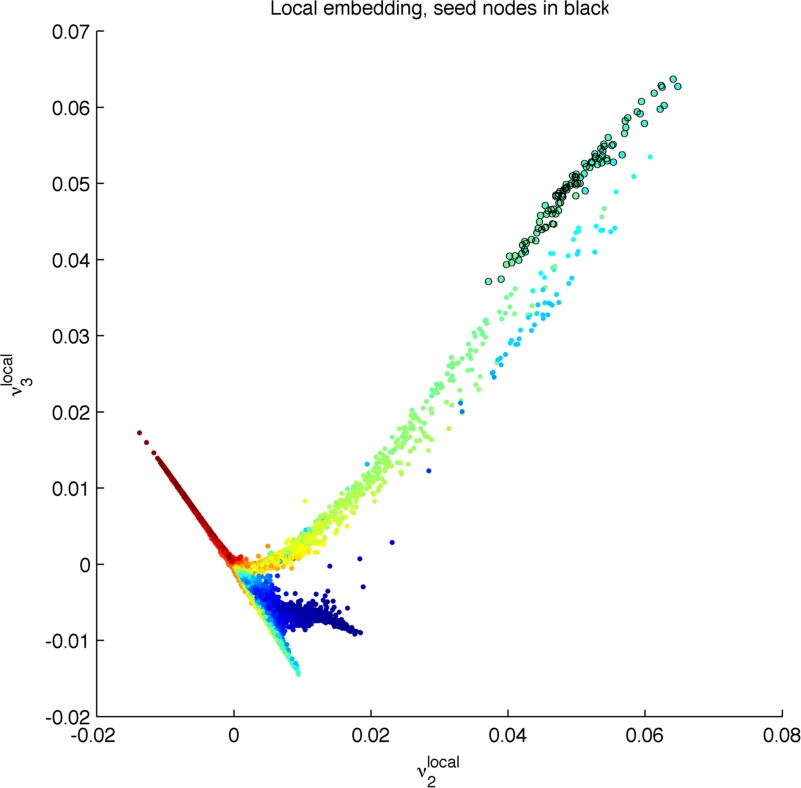}
	 \label{fig:effect-of-seed-B}
      } 
      \subfigure[]{
         \includegraphics[width=.30\textwidth]{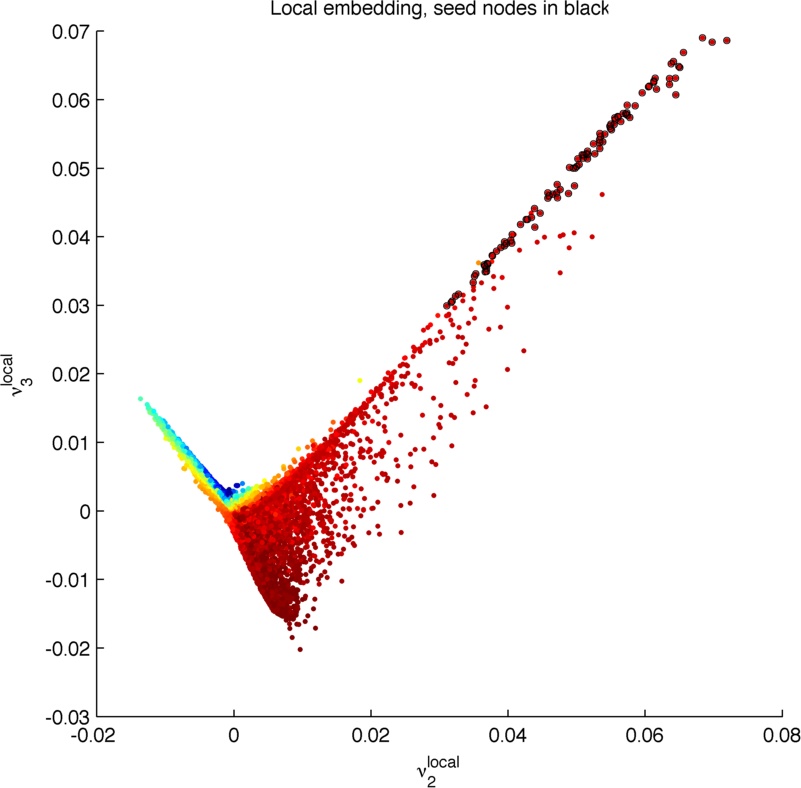}
	 \label{fig:effect-of-seed-C}
      } 
      \end{center}
\caption{(a) Embedding on global eigenvectors 2 and 3 of the lazy, autotuned Markov operator with $k=32$, with two seed node sets  highlighted in black. (b)--(c) Embedding on the leading two semi-supervised eigenvectors using seeds highlighted in (a), illustrating how semi-supervised eigenvectors can ``zoom in'' on particular parts of the data, down-weighting parts of the data far from the seed nodes. In all subfigures, points are color-coded by the second global eigenvector.}
\label{fig:effect-of-seed}	
\end{figure}

Here, we provide some intuition into the effect of the choice of seed vector $s$ on the local embedding introduced in Section~\ref{sec:methods}. 
For this purpose, we work with the lazy, autotuned Markov operator with bandwidth $k=32$, and we use the global embedding to provide meaningful local seeds. 
In Figure~\ref{fig:effect-of-seed-A}, we plot embedding of the galaxies on global eigenvectors 2 and 3, and color the points by the second global eigenvector. 
(This corresponds to Figure~\ref{fig:markov-densities-A}.)
Also indicated with black outlines are two subsets of galaxies.
These galaxies could be studied with global methods, as described in Section~\ref{sec:global}, but one might expect (and we have confirmed) that identifying such small-scale local structure with global eigenvectors---even if we permit the flexibility of multiple scales, multiple number of nearest neighbors, etc.---is extremely difficult and fragile.
Instead, these galaxies can be used as seed vectors for a local embedding with our main method.  
In each case, the seed vector used in \textsc{Generalized LocalSpectral} weas taken to be the indicator vector of the subset identified in Figure~\ref{fig:effect-of-seed-A}, and the correlation parameter $\kappa$ was set to 1/4 for each eigenvector. 
The seed sets were defined by manually choosing one data point of interest and taking its 100 nearest neighbors in the global embedding space.

In Figures~\ref{fig:effect-of-seed-B} and~\ref{fig:effect-of-seed-C}, we plot the resulting local embeddings.
That is, all the galaxy data points are plotted in the space spanned by the top two semi-supervised eigenvectors.
For comparison with Figure~\ref{fig:effect-of-seed-A}, we color the data points by the second global eigenvector (and the data points comprising the seed vector are indicated by black outlines). 
In both Figure~\ref{fig:effect-of-seed-B} and~\ref{fig:effect-of-seed-C}, galaxies similar to those in the seed vector are drawn away from the bulk of the embedding, offering a ``zoomed-in" view of the local region of interest defined by the seed vector; and galaxies very dissimilar to those in the seed vector  are given lower importance and clumped together. 
As with previous work that employed semi-supervised eigenvectors~\cite{HM14_JRNL}, these locally-biased embeddings can be used to visualize the data ``near'' the seed set of nodes, and they can be used to define features for use in downstream classification tasks.

\subsection{Locally-biased Learning}
\label{sec:local-learn}

To illustrate how the locally-biased embeddings can be used to perform improved locally-biased classification, we constructed a set of five local embeddings, one per class, seeded with a number of positive and negative examples of one class. 
That is, for a class $c$ and number of examples $n$, we choose $n$ random spectra of class $c$ and set the corresponding elements of $s$ to $+1$. 
We further choose $n$ random spectra from classes other than $c$ and set the corresponding elements of $s$ to $-1$. 
(We omit unclassified spectra from our sample, but we discuss these unclassified spectra below.) 
We expect that the local embedding with such a seed vector will better separate the class $c$ from the remaining spectra. 
For this study, we varied $n$ in the range $\{10,20,40,\ldots,640\}$, but for the sake of brevity we only present results with $n$=640. 
We calculate the top $9$ local eigenvectors for each class, with uniform correlation parameter $\kappa=1/9$. 
(Varying these parameters could potentially lead to improvements, as has been shown previously.)

We then trained a five-class logistic regression model using our global and local embeddings as features. 
We grouped features together into sets of five, taking a fixed number of local eigenvectors per class. 
Thus, we tested one model using the top local eigenvector from each class, another using the top two local eigenvectors from each class, and so on. 
We compared these with models built using the name number of global eigenvectors, that is, the top 5, 10, and so on. 
For each number of features, we cross-validated the model using 10 folds with a 10\%/90\% train/test split. 
The choice to train on the smaller side of the fold was made due to computational considerations, and we did not see a significant degradation in accuracy when the training and testing sets were reversed. 

The logistic regression model returns a vector of probabilities that a spectrum in the test set belongs to a given class. 
Using these probabilities, we constructed what we term ``probability confusion matrices", whose $i,j$ entry represents the average probability (over the test set) that a spectrum of type $i$ is assigned by the model to type $j$. 
Here, we take the class labels described in Section~\ref{subsec:embedding-class} as ground truth (although it should be noted that these labels are themselves imperfect, as they are the output of a separate model based on emission line ratios \cite{brinchmann04}).

\begin{figure}
      \begin{center}
      \subfigure[]{
         \includegraphics[width=.30\textwidth]{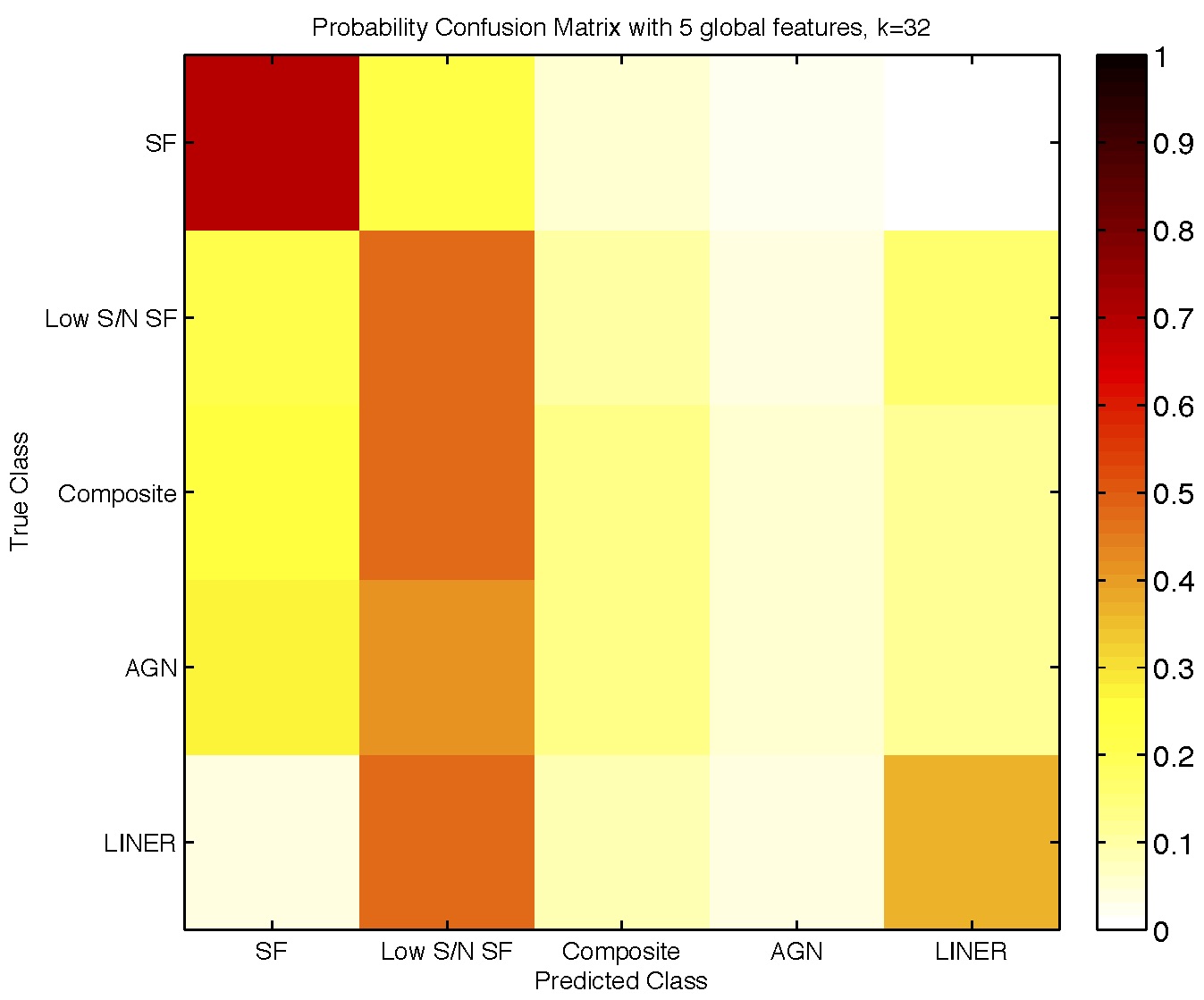}
         \label{fig:confmtxs-A}
      } 
      \subfigure[]{
         \includegraphics[width=.30\textwidth]{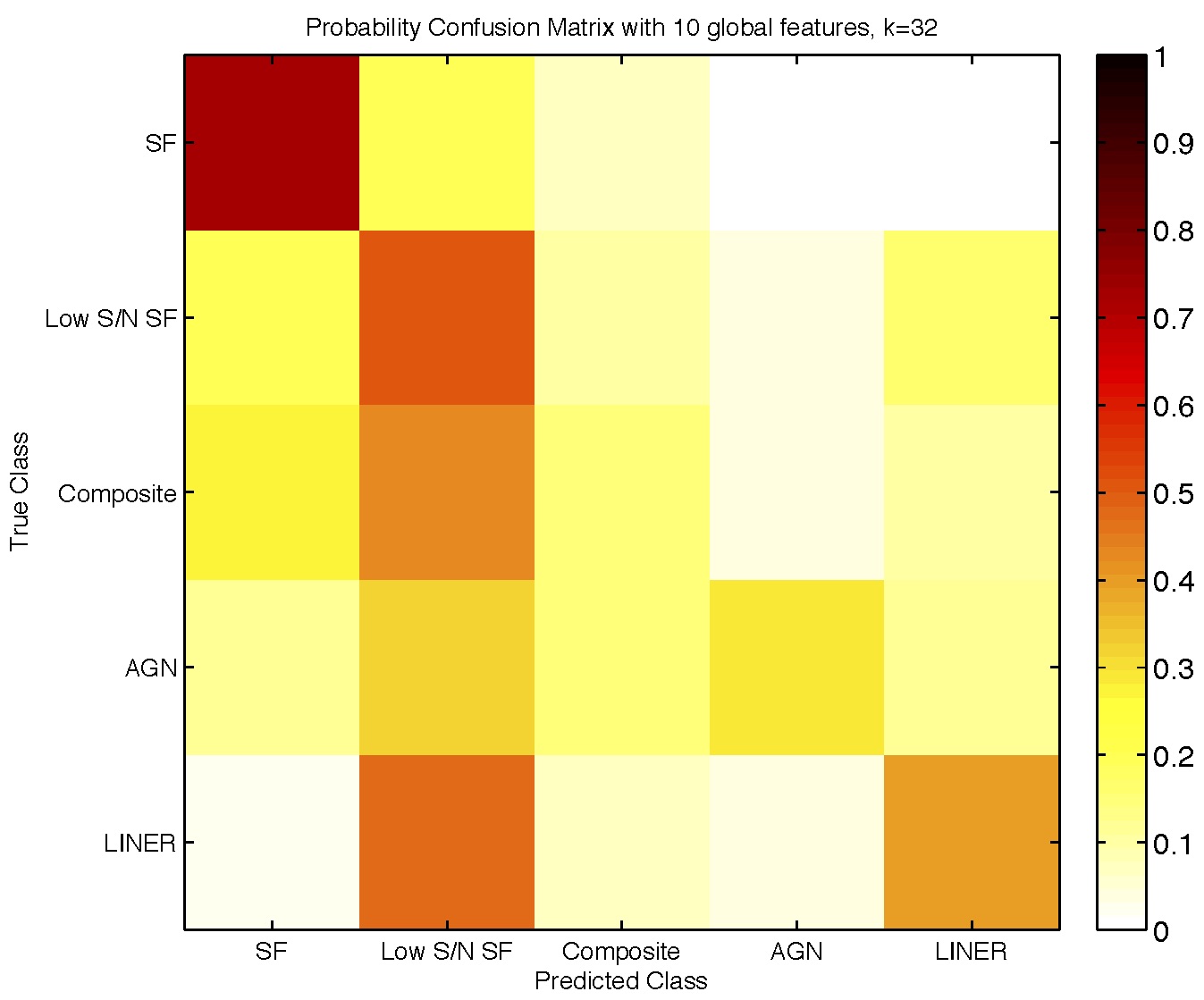}
         \label{fig:confmtxs-B}
      } 
      \subfigure[]{
         \includegraphics[width=.30\textwidth]{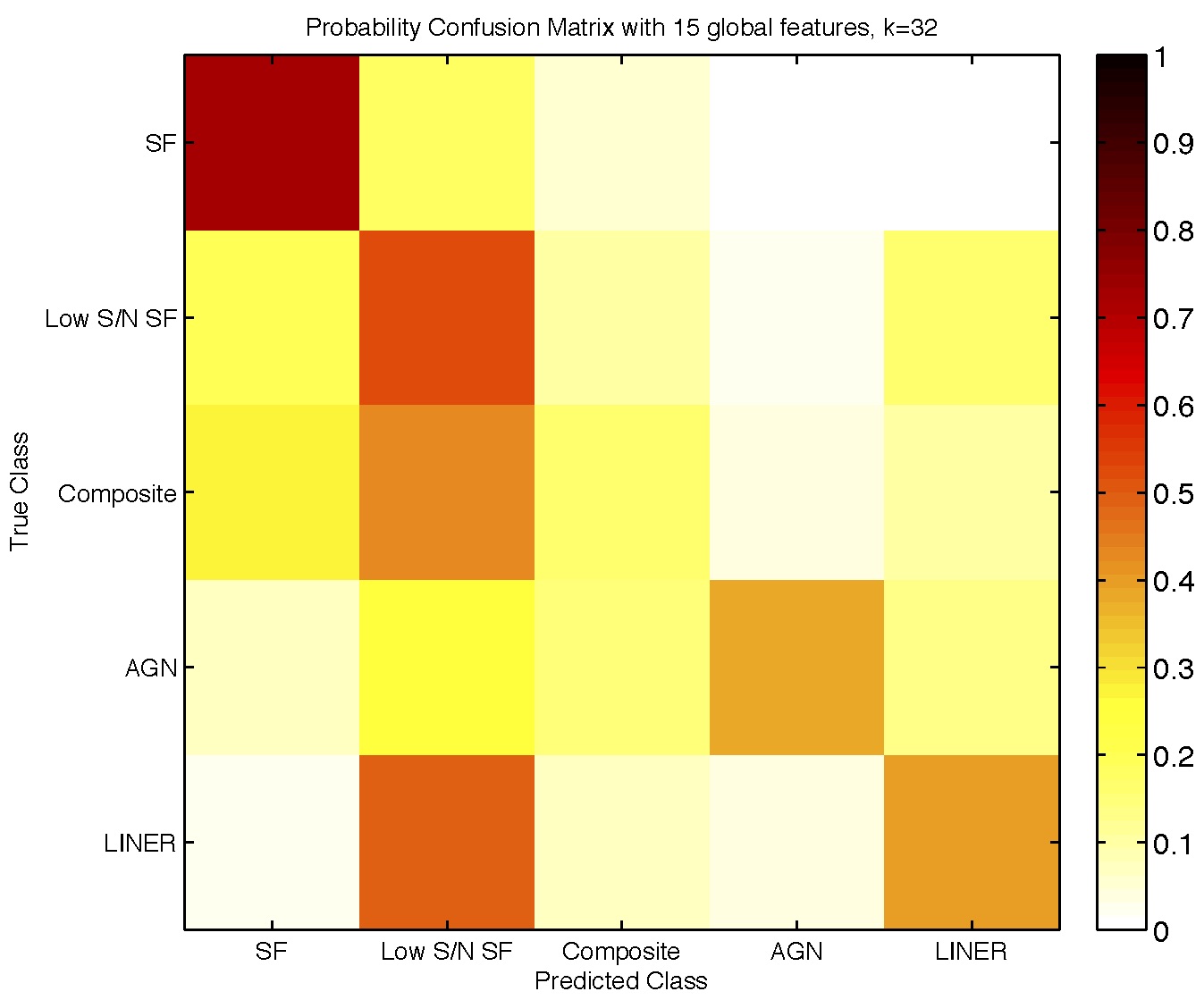}
         \label{fig:confmtxs-C}
      } 
      \subfigure[]{
         \includegraphics[width=.30\textwidth]{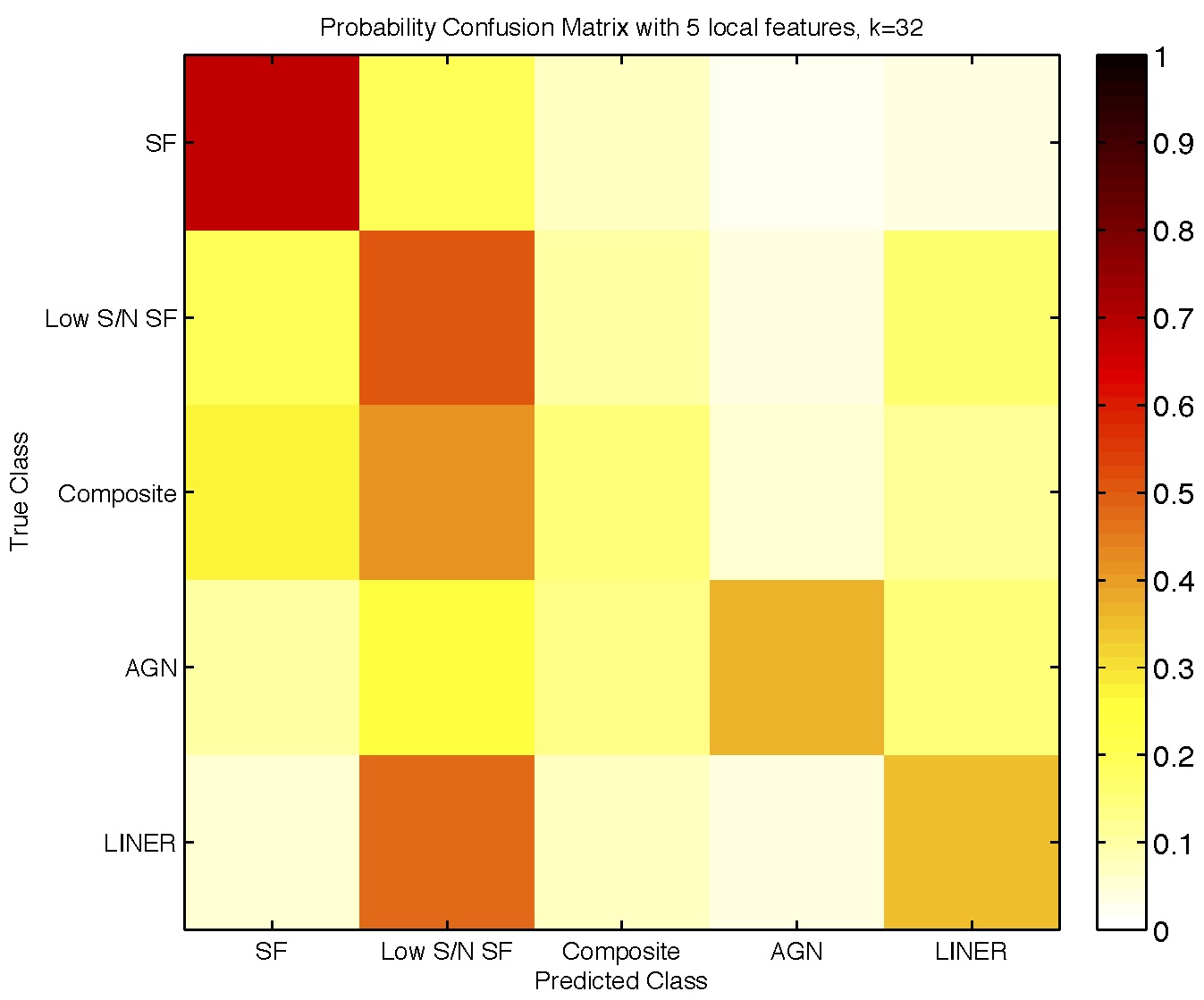}
         \label{fig:confmtxs-D}
      } 
      \subfigure[]{
         \includegraphics[width=.30\textwidth]{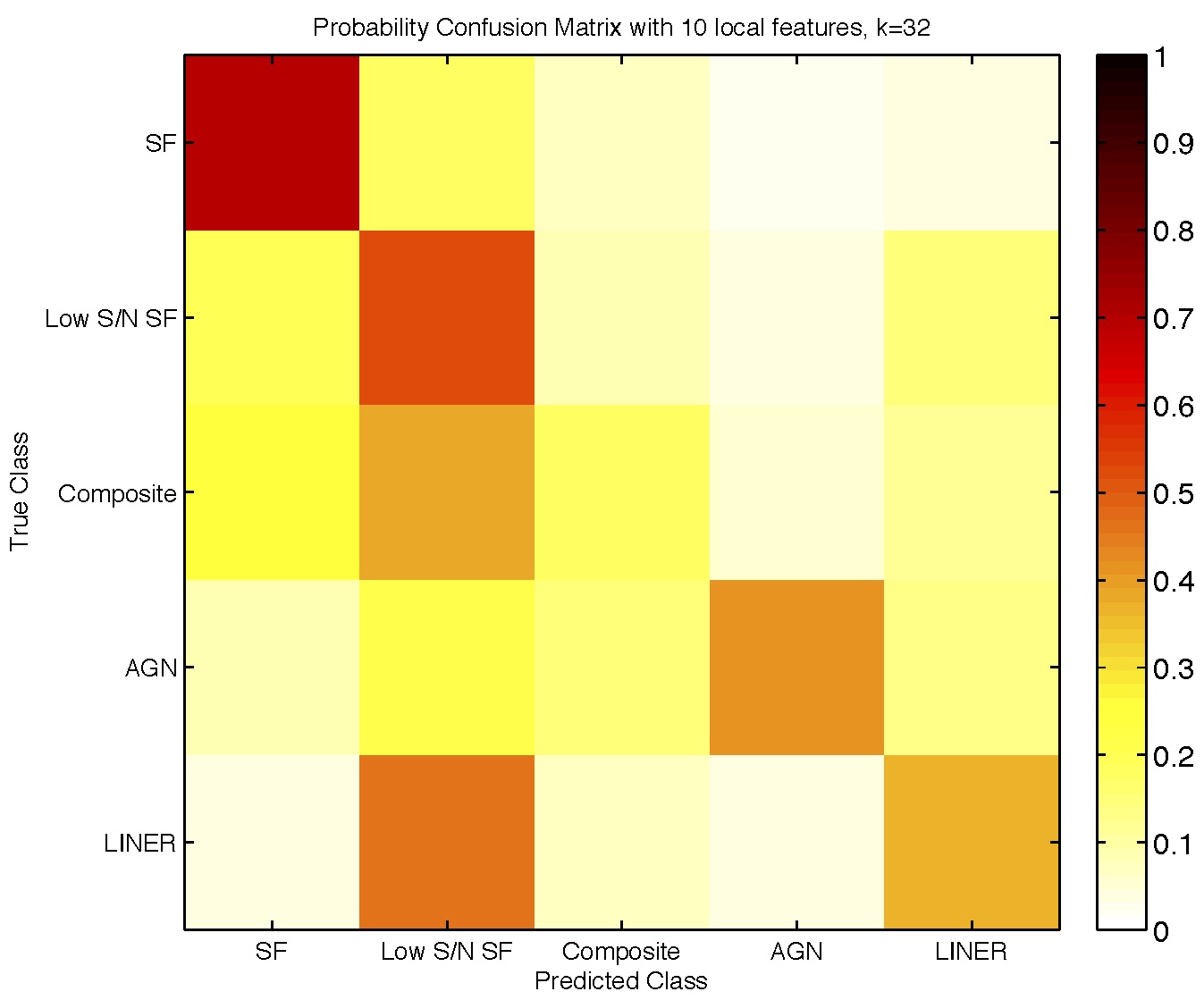}
         \label{fig:confmtxs-E}
      } 
      \subfigure[]{
         \includegraphics[width=.30\textwidth]{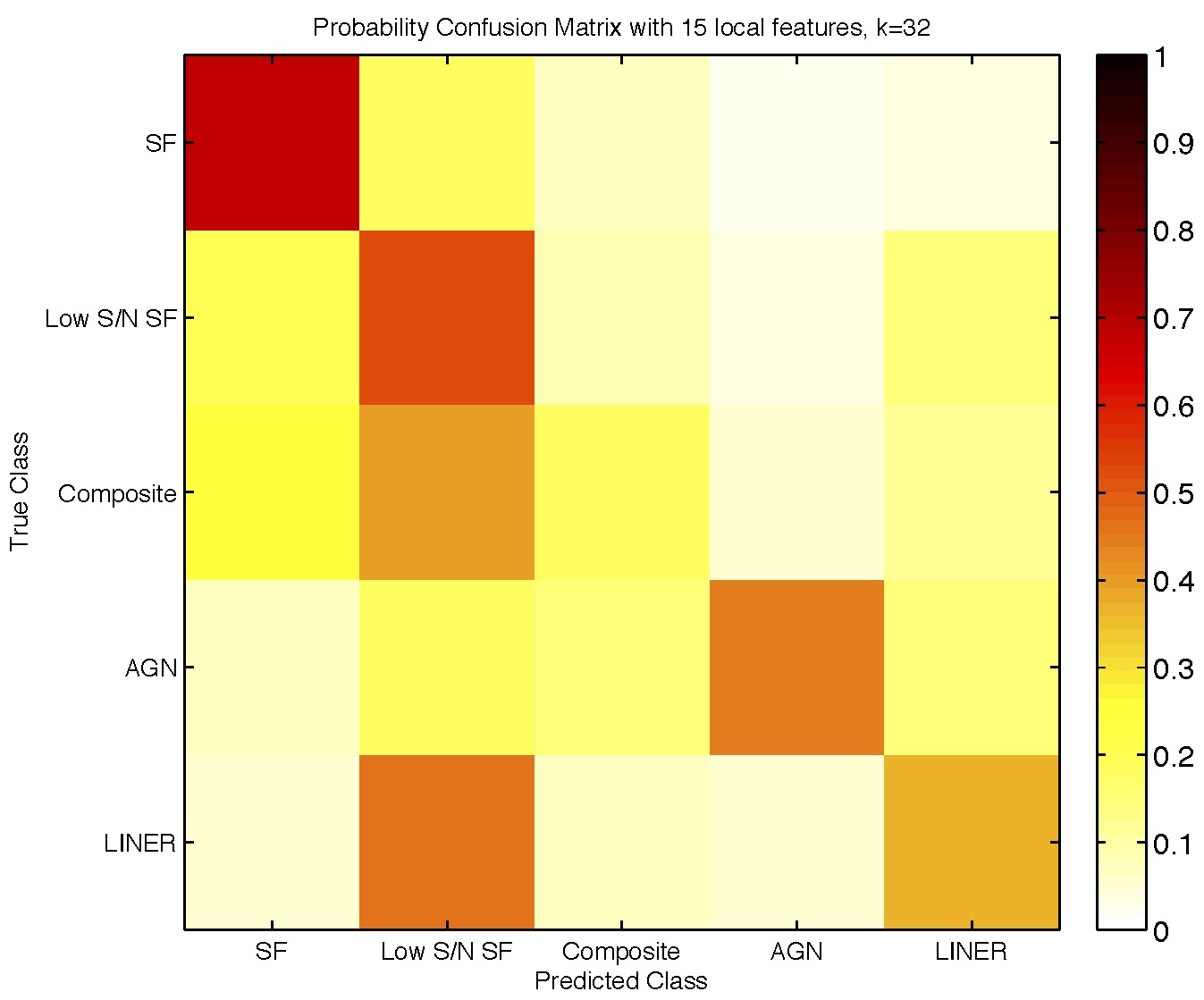}
         \label{fig:confmtxs-F}
      } 
      \end{center}
\caption{(a)--(c) Probability confusion matrices for five-class logistic regression using 5, 10, and 15 global eigenvectors of the $k=32$ graph; (d)--(f) using 5, 10, 15 local eigenvectors, with seed consisting of 640 positive and negative examples for each class. Results are cross-validated with 10 folds of 10/90 train/test splits.}
\label{fig:confmtxs}
\end{figure}

In Figure~\ref{fig:confmtxs}, we present the probability confusion matrices for models constructed with 5, 10, and 15 global and local eigenvectors of the lazy, autotuned Markov operator with $k=32$. 
In this figure, the top row corresponds to increasing number of global eigenvectors and the bottom to local eigenvectors. 
Comparing Figures~\ref{fig:confmtxs-A} and~\ref{fig:confmtxs-C}, we note that with 5 global features, the model has difficulty correctly classifying AGNs, while using 5 local features leads to a significant increase in the correct classification of such spectra. 
Thus, for a fixed budget of features the local model outperforms the global version, as expected. 
Increasing the number of features, global or local, leads to improved performance on this rare galaxy type, and the local model consistently outperforms its global counterpart.
(Clearly, the benefit of using local features is much less if we are interested in predicting very non-rare classes, as we have checked and confirmed.)

\begin{figure}
      \begin{center}
      \subfigure[]{
         \includegraphics[width=.44\textwidth]{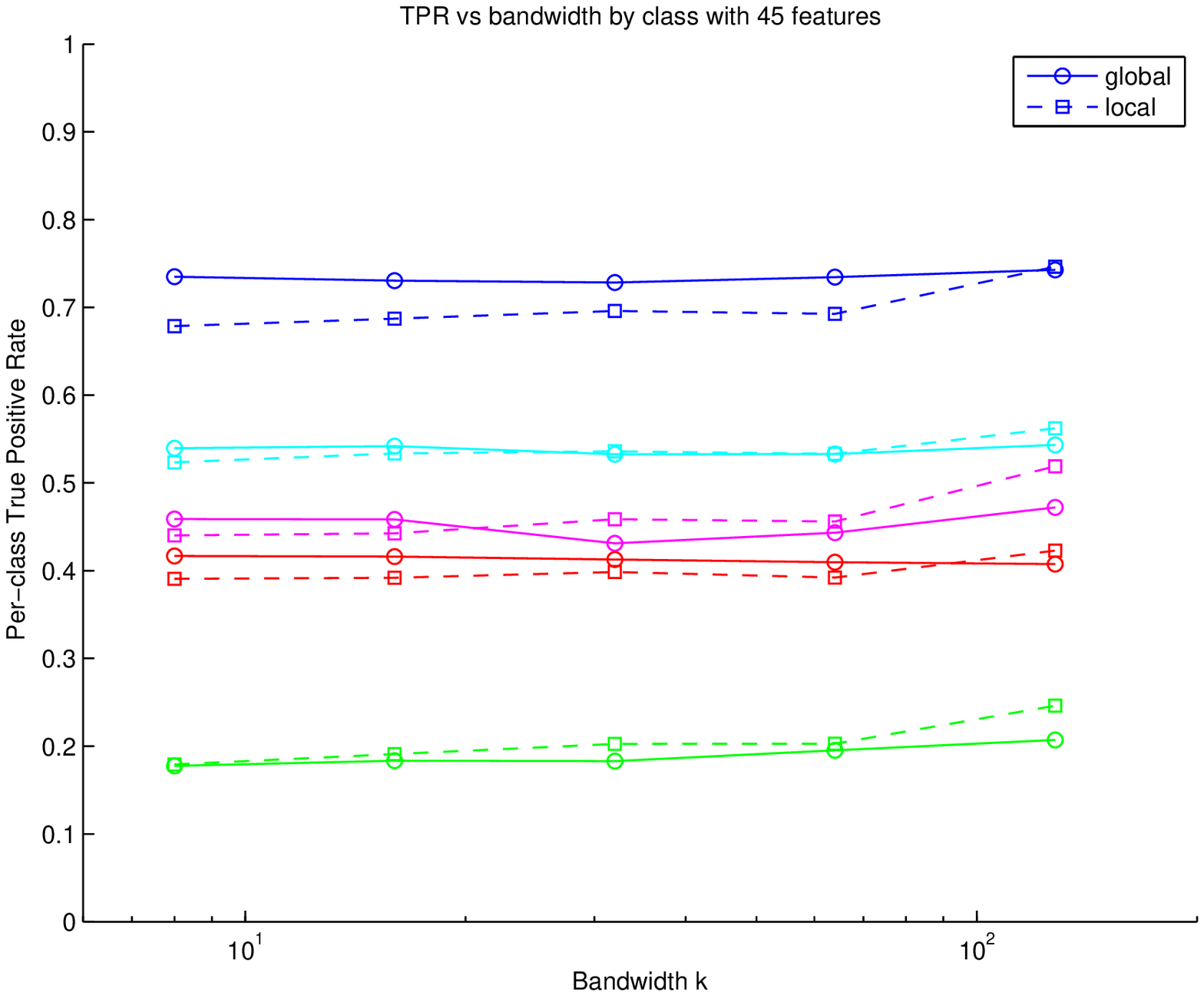}
         \label{fig:tpr-fdr-A}
      } 
      \subfigure[]{
         \includegraphics[width=.44\textwidth]{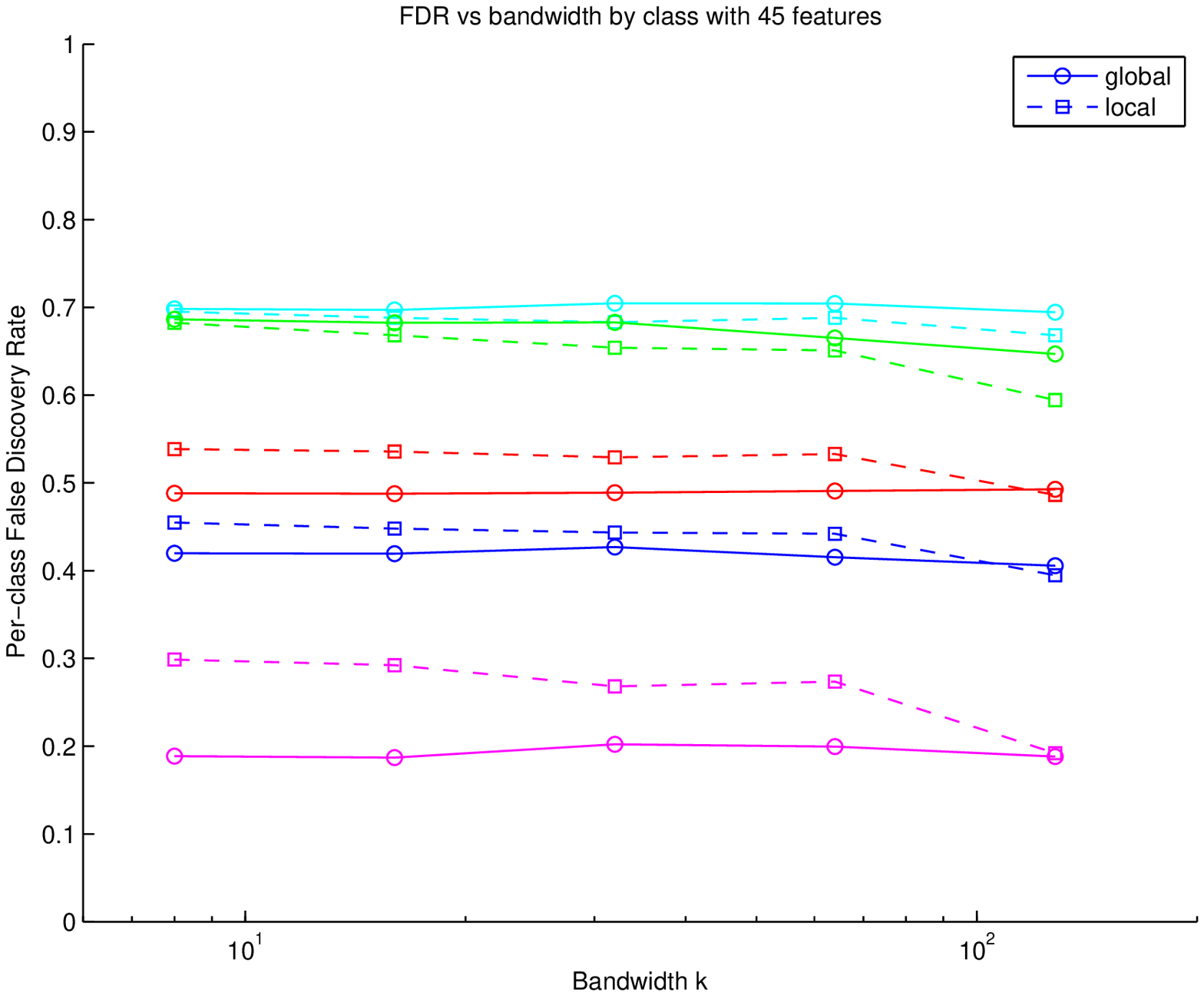}
         \label{fig:tpr-fdr-B}
      } 
      \end{center}
\caption{(a) Per-class true positive rates with 45 global and local eigenvectors as a function of bandwidth $k$; (b) per-class false discovery rates as a function of $k$.}
\label{fig:tpr-fdr}
\end{figure}

We also note that for all models shown, there is a large ambiguity between low signal-to-noise star-forming galaxies and LINERs. 
Indeed, for all cases the probability that a galaxy labeled LINER is classified as low signal-to-noise star-forming is greater than the probability that it is correctly labeled. 
This is not surprising in light of Figure~\ref{fig:global-class-label-C}, where significant mixing of these two types, colored cyan and red respectively, is visually evident. 
Finally, we tested the performance of the models as a function of the number of nearest neighbors $k$ in the underlying graph. 
In Figure~\ref{fig:tpr-fdr} we present the per-class true positive rates and false discovery rates as a function of $k$. 
Somewhat surprisingly, there is little dependence in either until $k=128$, in which case we see a slight improvement in performance. 

\subsection{Labeling Unclassified Spectra}
\label{sec:local-unknown}

For completeness, we include here a few comments about a use case that might have already occurred to the reader.  
There are a large number of galaxies for which there do not exist reliable ground truth labels into one of the classes we have been working with, and one might hope to define features---either global or locally-biased---to help with the classification of such unclassified spectra.
For example, given what we have observed above, there is a gradual transition between different classes, and the use of locally-biased features might define a locally-biased geometry to help with this imputation task.
We have examined this problem in some detail, and while this topic no doubt merits further attention, we have discovered that this is very challenging.
The reason is that unclassified spectra (even when they are not obvious outliers casued by experimental artifacts) typically have properties that are \emph{very} different than classified spectra.  
That is, they are unclassified for a good reason, i.e., since they are very different than spectra in one of the main classes, and in some sense they form their own (diverse) ``other'' cluster, whether viewed from a global or a locally-biased~perspective. 


\section{Conclusion}
\label{sec:conclusion}

We have presented a novel technique based on locally-biased semi-supervised eigenvectors to gain insight into the similarity of galaxies, and we have demonstrated the recovered nonlinear structure on the Main Galaxy Sample of the Sloan Digital Sky Survey. 
By constructing low-dimensional embeddings which respect local connectivity, we are able to visualize a range of astronomical phenomenon, e.g., the process of stellar evolution from hot, blue galaxies to cool, red ones. 
Unlike previously-used methods such as PCA or other recently-popular nonlinear dimensionality reduction methods, our method can focus on and highlight in a refined way local properties and/or global properties of the data.
Depending on the choice of knobs of the method, the embedding maps we create contain all galaxies, and they can be used to identify either large-scale global structure in the data or small-scale local structure in the data.

The main parameters that we empirically derive clearly correspond to changes in the continuum shape and the strengths of spectral lines, and these help disambiguate traditional BPT plots and isolated AGNs.
We observe that there are no disjoint groups of galaxies to indicate natural classes, but instead there are smooth continuous transitions between classes; and we observe that in many cases outliers (sometimes artifacts) are quite different than any other spectra in the data set.
Moreover, the locally-biased versions of these embeddings (which use semi-supervised seed labels to construct semi-supervised eigenvectors) demonstrate that the method can be used to enhance the embeddings such that we can better focus on galaxies of interest, e.g., rare galaxy types that can be overwhelmed by global~methods. 

To summarize, our method of locally-biased semi-supervised eigenvectors may be viewed as a new type of \emph{computational microscope} for astronomical data.
It can not only reproduce known properties of the data and identify outliers and artifacts, but it can also be used to enhance subtle trends around selected galaxies to facilitate new discoveries.
Obvious future work should focus on applying this new methodology to improve galaxy classification, to derive continuous spectral models, e.g., for photometric redshift estimators, and to understand better galaxy evolution with the help of stellar population synthesis models.

\section*{Acknowledgments}
We would like to thank Ilse Ipsen for valuable discussions.
This material was based upon work partially supported by the National Science Foundation under Grant DMS-1127914 to the Statistical and Applied Mathematical Sciences Institute. Any opinions, findings, and conclusions or recommendations expressed in this material are those of the authors and do not necessarily reflect the views of the National Science Foundation. MWM would also like to acknowledge the Defense Advanced Research Projects Agency and the Department of Energy for providing partial support for this work.



\newcommand{\etalchar}[1]{$^{#1}$}
\providecommand{\bysame}{\leavevmode\hbox to3em{\hrulefill}\thinspace}
\providecommand{\MR}{\relax\ifhmode\unskip\space\fi MR }
\providecommand{\MRhref}[2]{%
  \href{http://www.ams.org/mathscinet-getitem?mr=#1}{#2}
}
\providecommand{\href}[2]{#2}

%

\end{document}

%% file: artsettings.tex
\usepackage{amsmath,amssymb,amsthm,bm}
\usepackage{graphicx}
\usepackage{hyperref}
\usepackage{algorithm,algorithmic}
\usepackage[hang,small,bf]{caption}
\renewcommand{\phi}{\varphi}
\newcommand{\eps}{\varepsilon}
\newcommand{\R}{\mathbb{R}}

\theoremstyle{remark}

